%% file: 00_main.tex
\documentclass[sigconf]{acmart}
\AtBeginDocument{%
  }

\copyrightyear{2026}
\acmYear{2026}
\setcopyright{cc}
\setcctype{by}
\acmConference[UIST '26]{The 39th Annual ACM Symposium on User Interface Software and Technology}{November 02--05, 2026}{Detroit, MI, USA}
\acmBooktitle{The 39th Annual ACM Symposium on User Interface Software and Technology (UIST '26), November 02--05, 2026, Detroit, MI, USA}
\acmDOI{10.1145/3830398.3830627}
\acmISBN{979-8-4007-2856-3/2026/11}

\usepackage{minted}
\usepackage[most]{tcolorbox}
\usepackage{fancyvrb}
\definecolor{notebg}{HTML}{FDF3E7}

\usepackage{xspace}
\usepackage{xcolor}
\usepackage{enumitem}
\newcommand{\rv}{\textit{revibe}\xspace}
\newcommand{\rvb}{\textit{revibeability}\xspace}
\newcommand{\rvg}{\textit{revibing}\xspace}

\usepackage{xcolor}
\usepackage{balance}
\definecolor{LightGray}{HTML}{FFF5EE}

\usepackage{listings}
\begin{document}

\title{Revibing Code from Papers: Reimplementing HCI Artifacts}

\author{Eytan Adar}
\email{eadar@umich.edu}
\affiliation{%
  \institution{University of Michigan, Ann Arbor}
  \country{USA}
}

\author{Yoonjoo Lee}
\email{yoonjoo@umich.edu}
\affiliation{%
  \institution{University of Michigan, Ann Arbor}
  \country{USA}
}

\author{Ning-Er (Nina) Lei}
\email{nnlei@umich.edu}
\affiliation{%
  \institution{University of Michigan, Ann Arbor}
  \country{USA}
}

\author{Q. Vera Liao}
\email{veraliao@umich.edu}
\affiliation{%
  \institution{University of Michigan, Ann Arbor}
  \country{USA}
}

\author{Weirui Peng}
\email{weiruip@umich.edu}
\affiliation{%
  \institution{University of Michigan, Ann Arbor}
  \country{USA}
}

\renewcommand{\shortauthors}{Adar et al.}

\begin{abstract}
Software artifacts for most technical HCI research projects are unavailable. The lack of access to these imposes limits on academic knowledge production. It is difficult to: extend or reuse research artifacts; use strong baselines in evaluating follow-up work; and perform replication or reproducibility research. In this work, we demonstrate the potential of new agentic AI technologies to \rv interactive software: reimplement systems directly from research papers. To measure the success of the approach, we describe a \rvb metric. By \rvg recent research papers from UIST, and interviewing their original authors, we demonstrate the plausibility (and limitations) of revibed system. The results are encouraging. In many cases producing code suitable for strong baseline use. We argue that this may represent a fundamental shift in how we produce, use, and evaluate research artifacts in the technical HCI community.
\end{abstract}



\begin{teaserfigure}
 \includegraphics[width=\textwidth]{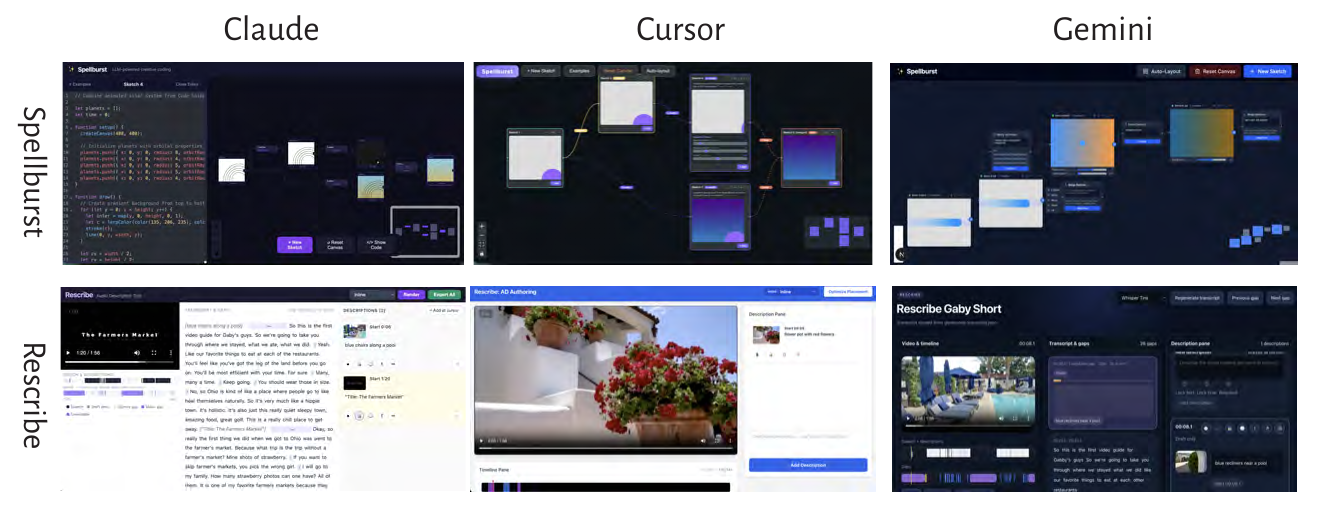}
  \caption{Screenshots of two revibed interfaces (\textit{Spellburst}~\cite{spellburst} and \textit{Rescribe}~\cite{rescribe}) produced by three different agentic environments (Claude Code, Cursor, and Gemini)}
 \Description{Screenshots of two revibed interfaces (Spellburst~\cite{spellburst} and Rescribe~\cite{rescribe}) produced by three different agentic environments (Claude Code, Cursor, and Gemini)}
  \label{fig:screenshots}
\end{teaserfigure}

\maketitle

\input{01_intro}
\input{02_related}

\input{03_measuring}
\input{04_method}
\input{05_experiment}

\input{06_results}

\input{07_discussion}

\input{08_conclusions}
\input{98_ack}

\bibliographystyle{ACM-Reference-Format}
\bibliography{02_x_bibliograph}

\appendix

\input{99_appendix}

\end{document}

%% file: 01_intro.tex
\section{Introduction}

While artifacts are the key aspect of technical HCI research, they are rarely made available. This limitation means that scientific knowledge production is constrained. The challenge in obtaining or reimplementing artifacts means that a researcher cannot easily extend or reuse technical innovations. Replication or reproducibility studies are impossible to implement. In the context of UIST-type evaluations, new tools cannot be properly evaluated through A-B studies because we do not have access to B, the \textit{strong baseline}. Arguably, the lack of access to artifacts skews knowledge production towards isolated novelty. That is, since it is harder to claim a contribution when a related system exists---but cannot be directly compared against---researchers might prefer to build in a completely new space. Today, there is very little to motivate the sharing of original code in technical HCI. There is perhaps even less to motivate reimplementation. Doing so is challenging and often unrewarding/unrewarded. Creating one artifact was hard enough, let alone trying to replicate another. 

Agentic programming tools~\cite{ding2025nl2repo,ge2025survey,lau2025design} may be changing this landscape. LLM-driven tools such as Microsoft's Copilot, Google's Antigravity, Cursor, Anthropic's Claude Code, and OpenAI's Codex~\footnote{see: \url{https://copilot.microsoft.com/}, \url{https://antigravity.google/}, \url{https://cursor.com/}, \url{https://claude.com/product/claude-code}, and \url{https://openai.com/codex/}}, have become the `habitat' for many developers and researchers. The consequence is that it is now far easier to \textit{create} certain kinds of technical artifacts. Additionally, it is also far easier to \textit{reimplement} artifacts. Agentic tools allow researchers to effectively reconstruct research artifacts when they are not available. 

However, the success of reimplementation will be influenced by several factors: namely, the content of the paper, the capabilities of the agentic tools, and the programming skills of the developer (i.e., the person carrying out the reimplementation). Vibe-coding, a narrower agentic programming paradigm introduced by Andrej Karpathy~\cite{vibedef}, assumes that a system can be built without the developer touching the code, but only through natural-language prompting. In this paper, we argue for the idea of using vibe-coding, or more specifically \rv-coding, to understand how well a system can be reimplemented directly from a technical HCI paper by a diverse set of agentic tools. By removing the influence of the developer's programming skills, we can identify a reimplementation `floor.' That is, we can determine how close we can get to a reimplementation based solely on the content of a paper.

In our definition, to \rv (the verb) is to reimplement a software artifact based only on the academic article describing it. A \rv (noun) is the reconstructed artifact. The success of a \rv will depend on what it is to be used for (in much the same way as any reimplementation). For example, if we are trying to replicate an experiment or reproduce a result, high fidelity may be necessary. If we are trying to extend the system or use some features of it, we may settle for lower fidelity. In this work, we focus on a middle ground: a sufficient reimplementation of that artifact so that it can serve as a \textit{strong baseline}. That is, we would like the reimplementation to be sufficiently interaction-- or feature--complete so that it can be used in an A-B style test. We propose a specific measure of success, \rvb, which uses a structured, system-specific testing rubric to elicit grading judgments as a quality measure. Neither \rvg nor \rvb requires that the developer touch any code directly.

To test this concept, we \rv ten different systems from recent UIST conferences (see Figure~\ref{fig:screenshots} for examples). These are software tools that do not have a publicly available implementation. For each of the papers, we \rv them 2-3 times each using Cursor, Gemini, and Claude (a total of 27 reimplementations). We show how \rvb rubrics can be automatically constructed and used to measure and \textit{improve} the revibes. Feeding the graded rubric back to the agent with instructions to fix the system can correct many system issues. With a fixed number of these feedback prompts ($N=2$), we demonstrate that we can achieve 94\% \rvb (mean, 80\% min, and 100\% max) on the tested systems. To further evaluate the revibes, we interviewed authors of three of the sampled papers. We also analyze different approaches to create \rvb rubrics and perform a qualitative evaluation of the successes and failures in the reimplemented systems.

Although imperfect, our results are encouraging. With improvements to agentic programming tools and more directed programmer feedback, we might expect even higher-fidelity revibes that can serve as strong baselines and other purposes of academic knowledge reproduction. The implications of \rvb of technical HCI papers are significant. In discussion, we describe the potential impact on system evaluation and how \rvg (and agentic development) can support new forms of knowledge construction in the technical HCI field\footnote{See \url{https://github.com/datamazelab/revibe-public} for additional supplementary information and scripts.}.

%% file: 02_related.tex
\section{Related Work}

We focus on two related lines of research: the use of AI in reproducing, replicating, or extending scientific work; and work on reproducibility and reimplementation in the HCI community.

\subsection{AI-Driven Replication}
LLMs, and in particular coding agents, have led to a significant body of research in replicating scientific work. However, this is largely in the context of data analysis or algorithmic/AI work. 

The potential of the approach is compelling for many disciplines. We have seen examples in political science~\cite{xu2026scaling}, social sciences~\cite{hu2025repro}, astrophysics~\cite{ye2025replicationbench}, natural language processing~\cite{yan2025lmr}, and machine learning~\cite{seo2025paper2code,hua2025researchcodebench}. Most of these projects can be evaluated directly by contrasting the results from the data analysis or algorithms done by the AI agents with what the original work produces. For example, if the AI created a modeling pipeline based on the paper and the model performance matches that of the original paper, the replication can be considered successful.

Various benchmarks have evolved to test (and drive improvement in) AI's capabilities to reimplement such research. For example, PaperBench is used to test LLM's capabilities in replicating AI research~\cite{starace2025paperbench}. Deep-Reproducer~\cite{chen2025deep} demonstrates an improvement on this benchmark with a multi-agent framework using the concept of `self-evolving debugging', achieving a 63.2\% replication score. We note that replication in this type of research is easier because evaluation is highly objective and can be automated. PaperBench, for example, uses an `LLM-as-judge' approach for evaluation. The output of technical HCI work is often software with an interface that cannot be evaluated in the same way.

Other related approaches use agentic tools to check the consistency between the paper and the released software~\cite{baumgartner2026scicoqa} or the reproducibility of key results~\cite{siegel2024core}. A key feature of many agentic systems is their ability to use `web tools' (i.e., access to search engines and retrieved results) to build software. For example, an agent may decide to use a specific web framework and will search for code examples or manuals. A more academic version of this is to build a database aligning code and techniques in papers~\cite{luo2025executable, zhao2025autoreproduce}. The agents can then isolate and reuse parts of the code that implement important functions. Such approaches can enhance scientific development by improving code~\cite{miao2025recode} and transforming code repositories into libraries for other use~\cite{jansen2025codedistiller}. Others have attempted to use LLMs to extend existing research with untested hypotheses~\cite{edwards2025rexbench} or to suggest new research ideas~\cite{si2025ideation}. However, these have had limited success with current models and approaches.

With \rvb, we are specifically interested in vibe-coded development directly from HCI papers. We do not count on the existence of the original code, data, or the availability of automated testing. Addressing these constraints requires the development of alternative ways to guide agentic tools and evaluate their results. We are interested in whether current popular agentic programming tools \textit{can} perform this task. Rather than just benchmarking their capabilities, we use a small but representative set of UIST papers ($N=10$) to also gain a qualitative understanding of how these tools succeed or fail in reimplementation. 

\subsection{HCI Reproducibility}
The ACM provides a number of `badges' to encourage open science~\cite{acmbadge}.  This includes: repeatability (same team, same experimental setup); 
reproducibility (different team, same experimental setup); and replicability (different team, different setup). Additionally, an `artifact available' badge (with sub-types of `reusable' and `functional') has also been produced. Different special interest groups and conferences have used these badges on papers.

However, HCI in general (and technical HCI specifically) has not adopted these badges (and corresponding goals). The short-lived RepliCHI movement attempted to create specific guidance around this for HCI~\cite{replichi11,replichi12,replichi13,replichi14}. Unfortunately, we continue to see a lack of transparency, replication work, and open science research in HCI~\cite{Ballou2021,hornbaek2014once,isenberg2024state,vanderdonckt2025context}.

Of particular importance is the lack of sharing of research artifacts~\cite{salehzadeh2023changes}, with some studies showing that only 2-3.5\% of authors share their code as open source in CHI~\cite{Echtler18}. Reasons for this include ``future research value, commercial value, and that they do not make sense outside the original context''~\cite{Wacharamanotham20}. Researchers are often reluctant to release code because it requires cleaning, documentation, and maintenance. This is demotivating, in particular, as the work is not often recognized.

Revibing provides a potential alternative route for the research community to gain access to the research artifacts without the original code or additional information. However, as we discuss below, there are opportunities for the original authors to support \rv with very little additional `lift.'

%% file: 03_measuring.tex
\section{Reimplementation and Revibeability}\label{sec:scoring}

\begin{figure*}[ht!]
    \centering
    \includegraphics[width=\linewidth]{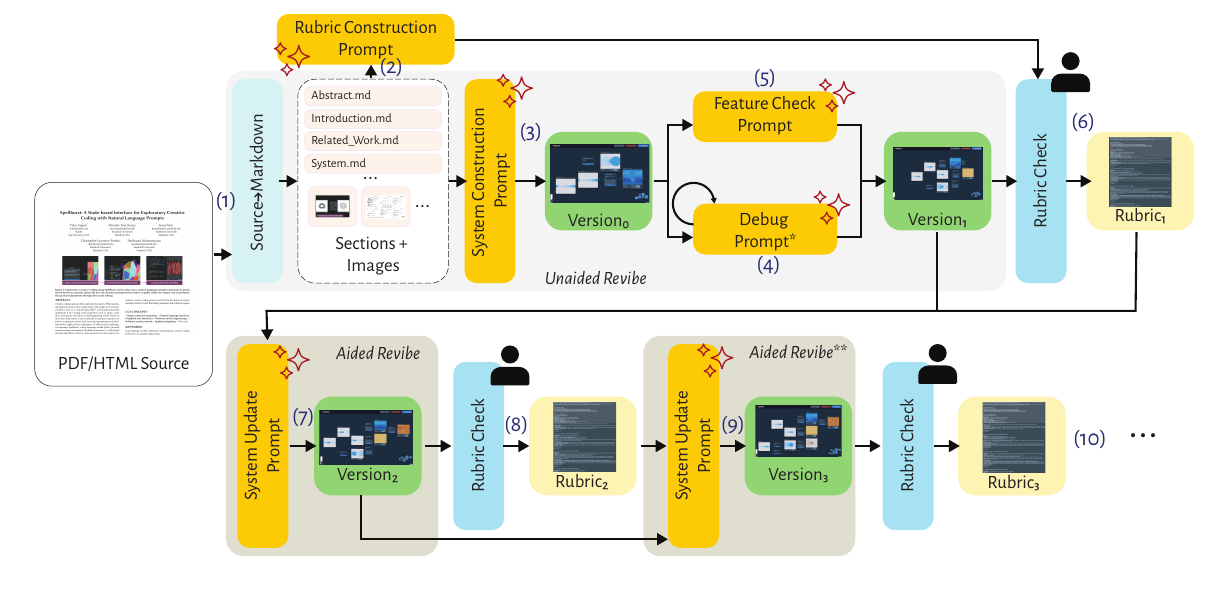}
    \caption{A figure of the \rv procedure implemented in this paper. The system `deconstructs' the paper into markdown and figures (1). From this, we construct a \rvb rubric for the paper using an automated prompt (2) and create an initial implementation (3). Debugging and feature checks (4 \& 5) produce a working implementation. This can then be graded (6) through the rubric. Repeatedly grading and asking the agentic system to update (steps 7-10) improves the \rv.}
    \Description{The figure describes the flow chart for the revibe process. A document is split into markdown files for each section with figures. This set of files is transferred to a Rubric construction prompt to produce a revibeability rubric. The files are also processed by a system construction prompt to produce version 0 of the system. A prompt to correct errors (debug prompt) is executed to ensure the initial version can run. Finally, a feature check prompt is executed. The result of this is version 1 of the system and represents the unaided prompt. The system is then checked against the rubric to produce a graded variant. The graded variant is passed with a system update prompt to fix implementation bugs. This is graded, and the result is version 2 of the system (and rubric 2). The process can be repeated multiple times to improve the system.}
    \label{fig:revibe}
\end{figure*}

Judging the success of a \rv will depend on the purpose of the reimplementation. If the goal is a replication study, we may not be satisfied with anything less than exact parity of all system features and interactions. If we are looking to extend one system with parts of another, we may only need to measure the quality of the parts we care about. In building a \rv for a strong baseline, a successful reimplementation should include all the key interactive features as described in the original work. Because what counts as key interactive features is system-specific, we propose a general test for \rvb based on a system-specific test rubric.

Specifically, for each system paper, we generate a test rubric (described fully in the Methods section below). The rubric is a form of QA testing for the described system. It specifies: the operations a user should perform (e.g., click on this, select that); what the user should expect to see; and what constitutes a failure, success, or something in between. A tester (e.g., the \rv developer in our case, but it can be another person or potentially automated) simply executes these test steps one at a time and reports on what they observe in the \rv.

We emphasize that our approach assumes the original source code is not accessible. This is both for practical (we target \rv to non-existing code) and experimental reasons (to ensure that the models have not been trained on existing repositories). The consequence is that we cannot directly apply automated test case generators to produce the rubric from the original code (e.g.~\cite{ramler2018adapting}). Instead, \rvb rubrics should be constructed from information available in the paper.

After each rubric test is graded as a failure, partial success, or full success, we can understand the \textit{disaggregated \rvb} of the revibed artifact. That is, which and how many of the features described in the rubric tests are successfully reimplemented (graphically depicted in figures such as Figure~\ref{fig:spellburst})? We define an \textit{aggregated \rvb} metric as:

\begin{center}
    $revibeability=\frac{w_s \cdot |t_{success}| + w_p \cdot |t_{partial}| + w_f \cdot |t_{failure}| }{w_s \cdot *|t_{all}|}$
\end{center}

Where $t_{all}$ are all the rubric tests, and $t_{success}$, $t_{partial}$, and $t_{failure}$ correspond to the tests that pass, partially pass, or fail (weighted at $w_s=2$, $w_p=1$, and $w_f=0$, respectively). This represents a simple but direct scoring rule that treats all tests within a class (failure, partial success, or success) the same. 

Alternative metrics may provide further weighting based on the importance of the tests. For example, one rubric item may represent a particularly critical feature of the system, and we may weight it higher. In addition, a more refined scoring scheme may be used for partial successes (e.g., 0.9, 1.1, etc.). We explore other variants in the Discussion section.

We observe a few implications for our choice of measure. First, \rvb may be unstable due to the stochastic nature of agentic programming tools. Different runs of the same \rv process on the same model \textit{may} yield different results. Additionally, as agentic tools integrate new prompts or models, their performance will vary over time. Moreover, different developers may score differently based on their interpretations of failure and success. In this work, we only use one \rv developer to grade and build each system once (per environment). Anecdotally, we found that our rubrics were largely unambiguous. Additionally, when experimenting with different prompts, we found that \rv outputs were largely consistent given the same agentic environment and model.

%% file: 04_method.tex
\section{Method}\label{sec:method}

The specific method we apply to \rv the code and to measure \rvb is illustrated in Figure~\ref{fig:revibe}. The four main steps for this process involve: extracting text and images from the source document (step 1 in the Figure); generating a rubric (step 2); performing an unaided \rv (steps 3-6); and performing one or more aided \rv procedures (steps 7-10). 

\subsection{Text Extraction}
Most modern models have fairly large context windows that can handle an entire academic paper. However, we found that we can reduce token costs and processing time by breaking a paper into chunks based on sections. This allows the agent to focus on sections that describe implementation and interaction details. 

We apply this process to HTML files (e.g., as produced by the ACM digital library) or to the published PDF. This module is a simple Python script. For PDF processing and image extraction, we use PyMuPDF~\footnote{\url{https://pymupdf.readthedocs.io/} and specifically the PyMuPDF4LLM sub-module for text extraction and Fitz for image extraction}. This procedure ensures that two-column PDFs are processed correctly. The output of this code (Figure~\ref{fig:revibe}-(1)) is a set of markdown files (with filenames corresponding to section headers) and images. These are not cleaned in any other way, so some spurious artifacts (e.g., headers and footers) might remain.

\subsection{Rubric Development}
The second step in the \rv pipeline is to create a system-specific scoring rubric to judge the success of the reimplementation (Figure~\ref{fig:revibe}-(2)). As we describe in Section~\ref{sec:scoring}, the rubric should include a set of tests that correspond to all the key interactive features or interactions described in the original artifact. Each test should specify the operations a tester should perform and the criteria for failure, success, or partial success.  For example, assume a paper indicates that clicking some button X will result in some event Y. A \rvb rubric test will first ask the tester to confirm that button X exists and then indicate what should happen when it is clicked. If the same event (Y) happens,  the tester can mark this test as a success. If nothing happens, they should mark this as a failure. The graded rubric can be used to measure the \rvb based on the aggregated metric or disaggregated scores, as described in Section~\ref{sec:scoring}. The graded rubric can also be provided back to the agentic programming tool for aided \rv to fix the failed features and improve the reimplementation (see below, Section~\ref{aided-rv}). 

To enable standardized testing, we constructed a prompt for generating a \rvb rubric automatically from the paper. This approach has practical implications as it normalizes the types of human input available in aided \rv to improve the reimplementation. Specifically, the input includes only direct feedback on whether the \rv can perform the intended functionalities and does not depend on the tester's own reading of the paper or their technical skills (vibe coding or otherwise). 

In practice, there may be different ways to create the rubric. In an ideal scenario, the authors of the original paper would modify the auto-generated rubric to ensure it aligns with their intended system functionalities. They could also label which tests are more important---for example, if they reflect the key contributions of the original artifact---and these tests will be weighted more heavily. Slightly less ideal would be to allow the person \rvg to modify the rubric based on their reading of the paper or understanding of the supplementary materials (e.g., a video of the system working). This assumes that the person will perform better than the LLM in either creating the rubric or identifying key features/tests.

We iterated on the \textit{Rubric Construction Prompt} (see step 2 in the Figure) to arrive at the following: 

\begin{tcolorbox}[
  colback=notebg,     
  colframe=notebg,    
  boxrule=0pt, arc=0pt,
  left=2pt, right=2pt, top=2pt, bottom=2pt,
  breakable]
\ttfamily\small
\begin{Verbatim}[fontsize=\small, breaklines=true, breakanywhere=true, breaksymbolleft={}]
Based on the contents of the paper and any images determine (see md and png/jpeg/jpg files) a list of key UX elements. Describe a list of manual tests that a human can use to validate the functionality. For example, if a key interaction is "loading" a file by clicking on an "open" button, the instructions might direct the user to navigate to where the button is, click open, and verify that the content was loaded (with instructions on what verification looks like). The instructions should provide a detailed enough rubric on what success, partial success, and failure look like. A human should be able to use this to "grade" how well an implementation meets the description of the system as intended by the authors. Save these instructions to a interaction_validation.md file. The format of the document should be:

Test Plan Structure
Tests are organized by application (main windows/screens), then cross-application integration, then edge cases. Each test includes:

ID: Unique identifier
Prerequisites: Required setup
Steps: Numbered actions
Expected Result: What should happen
Success Criteria: Three levels (Full Success / Partial Success / Failure) with descriptions for each
Observed result: A blank space where the user can report on success and provide any details on incomplete/incorrect behaviors.
\end{Verbatim}
\end{tcolorbox}

Note that we do not argue that this is the \textit{best} such prompt. What it allows for is the construction of evaluation criteria that can be uniformly applied across all revibed artifacts. Two example `tests' for the \textit{Spellburst} system~\cite{spellburst} are presented in Appendix~\ref{sec:rubric_example} (and in the web supplement). Although we evaluated the generation of rubrics with multiple models, we ultimately opted to use rubrics produced by Claude Code (\texttt{Sonnet 4.6}). This produced the most detailed rubric (e.g., 38 tests in one case produced by Claude versus 6 produced by Gemini). While this can potentially create some bias towards systems created by Claude, we ultimately did not see this in the results.

\subsection{Unaided Revibe}

To perform an Unaided Revibe (i.e., without specific guidance from the human), we provide the paper markdown/images to each of the coding tools with the following prompt (Figure~\ref{fig:revibe}-(3)):

\begin{tcolorbox}[
  colback=notebg,     
  colframe=notebg,    
  boxrule=0pt, arc=0pt,
  left=2pt, right=2pt, top=2pt, bottom=2pt,
  breakable]
\ttfamily\small
\begin{Verbatim}[fontsize=\small, breaklines=true, breakanywhere=true, breaksymbolleft={}]
You will be implementing the [SystemName] system. This is based on a paper. The paper's content is split into markdown files and images in the 'paper' directory.  Proceed in the following steps:

(1) Analyze the features of the system and user interface (use the paper content and images)

(2) Create a new implementation of the system that supports all the described features. Create this as a web app if you can, but you can use an alternative approach if the design calls for it (e.g., a Python backend). You may select alternative libraries or languages for implementation that deviate from those used in the paper.

(3) When you are done, create a DEVELOPER.md that provides a detailed walkthrough of what a developer needs to know and install to run the system. Also, create a README.md that describes how to launch and use the system.

Additional instructions:
* You may ask questions if something is unclear but provide alternative choices.
* If you need access to an API key, you can check for an available .env file. If the key does not exist, you can leave instructions in the .env file. Notify me, and I will fill it out.
* It is ok to use roughly equivalent APIs (e.g., if the API calls for using a OpenAI model and you only have access to Gemini, you can use that)
\end{Verbatim}
\end{tcolorbox}

The prompt was designed after some iterations (e.g., indicating that equivalent APIs were okay to use and asking the developer if something was unclear). We aimed to provide minimal guidance, allowing the agentic tool to make its own development decisions while also ensuring that agents do not get stuck on specific issues. The human protocol specified that we should answer agentic questions as well as possible with a shallow read of the paper and by going with suggested options when possible. As we discuss below, questions were largely restricted to the approval of steps (e.g., can I install these Python packages?) and not about the paper content. In some cases, the agent would split the task into steps, asking for approval as it went. For example, step 1 would be to implement a draft of the GUI and step 2 would be to improve it.

Note that, because our focus is on the reimplementation of software artifacts with a user interface, there are some inherent limitations to the prompt. For example, we ask the agent to produce a web app if possible. In some situations, a software artifact would work better as a library or as a plugin to an existing system (e.g., a Jupyter plugin in the case of mage~\cite{mage}). Because our selected papers did not require specific hardware or underlying tools (e.g., an Adobe Photoshop plugin), this prompt worked for our purposes. 

After the initial codebase is produced, we follow the generated instructions (in README.md) on how to run the code. Any observed error messages that prevent the software from running are fed back to the coding environment with the prompt (Figure~\ref{fig:revibe}-(4)) to produce $version_0$ of the system:

\begin{tcolorbox}[
  colback=notebg,     
  colframe=notebg,    
  boxrule=0pt, arc=0pt,
  left=2pt, right=2pt, top=2pt, bottom=2pt,
  breakable]
\ttfamily\small
\begin{Verbatim}[fontsize=\small, breaklines=true, breakanywhere=true, breaksymbolleft={}]
I have observed the following errors:
[error messages]
Please debug these and fix them.
\end{Verbatim}
\end{tcolorbox}

Most often, these were errors where the system did not install a specific package. In a few cases, we observed server or client errors (observed in the console or terminal window) that would prevent the system from running. This simple prompt allows the system to correct any obvious problems. We ran this prompt as many times as needed until the application launched (practically, most revisions required a maximum of 2-3 error fixes). No other feedback was provided about bugs. While we used command line versions of the systems, we expect that those agentic environments that can run and test the browser themselves (e.g., Google's Antigravity) might detect client errors automatically.

In piloting, we noticed that some agentic tools prioritized implementation momentum--making simplifying decisions. Their prompts include language such as: \textit{Avoid over-engineering. Only make changes that are directly requested or clearly necessary. Keep solutions simple and focused}, and \textit{Choose the most sensible approach and keep moving.}~\footnote{\url{https://github.com/Piebald-AI/claude-code-system-prompts}}. The result was that in some implementations, certain system features might be implemented using simpler algorithms or left to `future work.' To ensure a more complete \rv, we designed one final prompt to `complete' the implementation (Figure~\ref{fig:revibe}-(5)):

\begin{tcolorbox}[
  colback=notebg,     
  colframe=notebg,    
  boxrule=0pt, arc=0pt,
  left=2pt, right=2pt, top=2pt, bottom=2pt,
  breakable]
\ttfamily\small
\begin{Verbatim}[fontsize=\small, breaklines=true, breakanywhere=true, breaksymbolleft={}]
Compare the implemented features to the features described in the paper. Add any missing features.
\end{Verbatim}
\end{tcolorbox}

This produced a $version_1$ of the system that was then evaluated using the rubric. Each tester followed the rubric instructions to the best of their ability and graded whether the system fully succeeded, partially succeeded, or failed each of the rubric tests. Any specific observations about GUI behavior were noted in the document (see the web supplement for examples). The result of this was a first graded rubric, $rubric_1$, and an initial measure of \rvb (Figure~\ref{fig:revibe}-(6)).

\subsection{Aided Revibe}
\label{aided-rv}
The graded rubric was provided back to the agentic tool to perform an aided \rv. Clearly, proficient developers could provide further `aid'. However, as we note, this approach normalizes feedback and provides us with a \rvb floor. The specific prompt was:

\begin{tcolorbox}[
  colback=notebg,     
  colframe=notebg,    
  boxrule=0pt, arc=0pt,
  left=2pt, right=2pt, top=2pt, bottom=2pt,
  breakable]
\ttfamily\small
\begin{Verbatim}[fontsize=\small, breaklines=true, breakanywhere=true, breaksymbolleft={}]
I have completed an evaluation of the system and tested its features. Please fix any reported issues in the attached rubric. [@pointer to rubric file]
\end{Verbatim}
\end{tcolorbox}

The output of this step is a new version of the \rv (i.e., $version_2$, Figure~\ref{fig:revibe}-(7)) that could be evaluated again (Figure~\ref{fig:revibe}-(8)) to produce a new graded rubric ($rubric_2$). This process could optionally be repeated (Figure~\ref{fig:revibe}-(9,10)) to further improve the \rv. 

%% file: 05_experiment.tex
\section{Revibing UIST}
To test the \rvg protocol, we started with a set of popular software-focused papers from UIST from 2020 to 2024 (identified on Semantic Scholar based on citations). This time range was selected because: (a) the systems were around long enough to assess their popularity, (b) they were likely to use modern interface approaches, and (c) they were less likely to be vibe-coded themselves (though we cannot eliminate this possibility entirely).

We removed papers that had public implementations or required additional hardware (e.g., VR or fabrication-focused systems). This yielded 20 systems (see Appendix~\ref{sec:systems}). The full 20 were used for experiments on \rvb rubrics (see Section~\ref{sec:rubrics}). From these, we selected 10 for \rvg. These were selected roughly randomly. However, we found that many recent UIST papers leverage LLMs as a component of the system (e.g., Spellburst~\cite{spellburst}). To validate that \rvg works for non-LLM-centered systems we oversampled these. 

For each system, the authors (acting as assessors) used the approach described in Section~\ref{sec:method}. The assessors were a combination of PhD students in CS, postdocs or faculty, all with previous experience in technical HCI systems. While we believe this population to be representative of those likely to \rvb systems, we note that the execution and evaluation of the approach do not require any particular deep technical skills. While we did not perform a specific calibration, the approach for rubric grading was discussed among the team before revibing began. We utilized three agentic tools for this work: Cursor, Anthropic's Claude Code, and Google's Gemini (all with the command line versions). These systems are among the most popular at the time of writing (spring 2026), and we largely used the default models for each of the tools: a combination of Google's \texttt{gemini-3.1-pro} and \texttt{gemini-3-flash} for Gemini; Anthropic's \texttt{claude-sonnet-4-6} for Claude, and OpenAI's \texttt{gpt-5.4-xhigh} for Cursor~\footnote{For two systems (mage and Spellburst), we also tested Cursor with \texttt{claude-sonnet-4-6} (see Appendix~\ref{sec:cursor}).}.

Our protocol used one round of unaided \rv followed by \rvb grading. We then applied two rounds of aided \rv, producing intermediate and final \rvb grading. We note that in this initial study of \rvb, we opted for breadth rather than depth. That is, we did not continue the \rv process until \rvb scores `saturated.'  However, in most cases, we observed that the performance of pure \rv coding either hit the maximum value or became asymptotic.

In addition to the \rv experiments, we interviewed the authors of the Spellburst (one author), B2 (one author), and mage systems (three authors). This was approved by our institutional IRB and took place over Zoom in a 30-40 minute call. The authors were not compensated for this session. They were shown the different reimplementations and asked a number of questions (see the web supplement). We did not explicitly indicate at the start of the session that the systems were created through agentic programming (to avoid biasing the discussion). However, it is likely that the authors could infer what we had done from earlier parts of the interview. Regardless, we described the approach in a debrief at the end.  We integrate author comments into the qualitative analysis of \rv performance below.

%% file: 06_results.tex
\section{Revibeability Results}
\begin{figure}
    \centering
    \includegraphics[width=.8\linewidth]{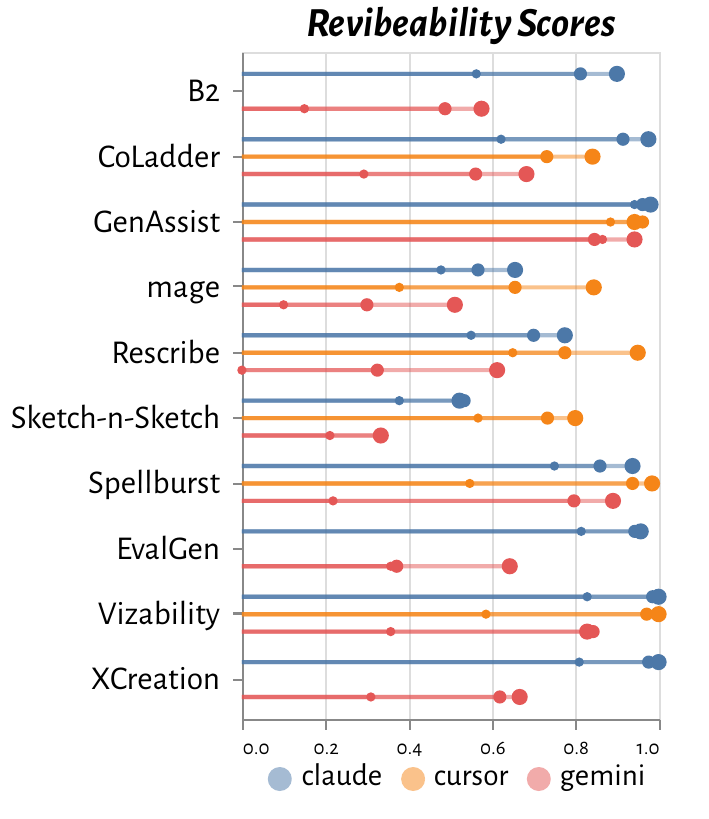}
    \caption{Summary of \rvb scores in our experiments (calculated as the aggregated \rvb score). The color of the lines/dots represents the vibe-coding environment. The size of the dots, from smallest to largest, indicates the \rvb score for each run (after the unaided \rv, post-aided \rv 1, post-aided \rv 2).}
    \Description{The figure reports the results of the revibe process for each system and agentic framework (27 lines in a `lollipop’ chart). The specific values of importance are available in the text. The figure demonstrates that most systems achieve revibeability scores above .8 within 2 aided steps. It also illustrates that most revibe procedures have an improved score as the aided revibe is applied. In general, we see that Claude and Cursor beat Gemini in revibeability.}
    \label{fig:revibeability}
\end{figure}

Figure~\ref{fig:revibeability} depicts the \rvb performance of our implementations. The dots on each line indicate the \rvb score (as calculated from the equation in Section~\ref{sec:method}) after each run (unaided, aided 1, and aided 2). All ten systems were revibed with Gemini and Claude. Seven were also revibed using Cursor. The cost of \rvg varied by system and model, but most automatic \rv loops ran in under 30 minutes. The first (unaided) \rv for most systems cost under \$15 with the price dropping for second revibes as code was being modified rather than created from scratch. The dominant time cost was in human assessment. The first iteration required 1-2 minutes per test. Subsequent assessments were faster, as tests that succeeded only required a quick confirmation.

Revibeability scores were monotonic over runs in almost all cases ($revibeability_{unaided} \leq revibeability_{aided1} \leq revibeability_{aided2}$). In one case, we saw a decrease from unaided to aided 1 (Gemini for GenAssist), and in three cases, we saw a decrease from aided 1 to aided 2 (GenAssist with Cursor, Sketch-n-Sketch with Claude, and Vizability with Gemini). All negative changes were under .02 (i.e., roughly the same \rvb).

The `best-run'---the highest score achieved by any agent after any run for each system---had a mean of 0.939 across all systems (min=0.8, max=1, stdev=0.069). Two systems (\textit{Vizability} and \textit{XCreation}) achieved a perfect \rvb score. Cursor-revibed systems had a mean \rvb of 0.912 (min=0.8, max=1, stdev=0.08). Claude revibes had a mean \rvb of 0.87 (min=0.53, max=1, stdev=0.16). Finally, Gemini performance \rvb had a mean of 0.67 (min=0.33, max=0.94, std=0.18).  

We find a correlation between the length of the \rvb rubric and the maximum score achieved (see Appendix~\ref{sec:rubrics}). With increased rubric complexity, there are naturally more points of possible failure. While it is likely that more complex systems will have longer rubrics, we leave this analysis for future work.

\begin{figure}[t]
    \centering
    \includegraphics[width=\linewidth]{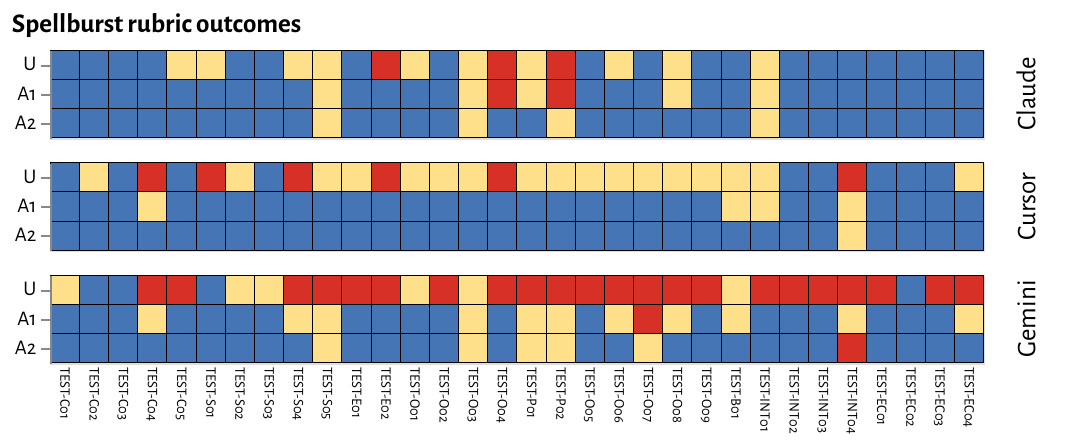}
    \caption{Revibeability performance for \textit{Spellburst}. Red/Yellow/Blue correspond to failure, partial success, and full success fo each test (respectively)}
    \Description{The figure is a heatmap for the Spellburst revibe. Each row is the rubric’s results after each run (an aided and two unaided). Each column is a test. The cell colors red, yellow, and blue correspond to failure, partial success, and complete success, respectively. System rows are grouped to show changes in score over iterations. In this case, Claude goes from .75 to .86 to .93. Cursor goes from .7 to .95 to 1.0. Gemini goes from .22 to .8 to .89.}
    \label{fig:spellburst}
\end{figure}

\begin{figure}[t]
    \centering
    \includegraphics[width=\linewidth]{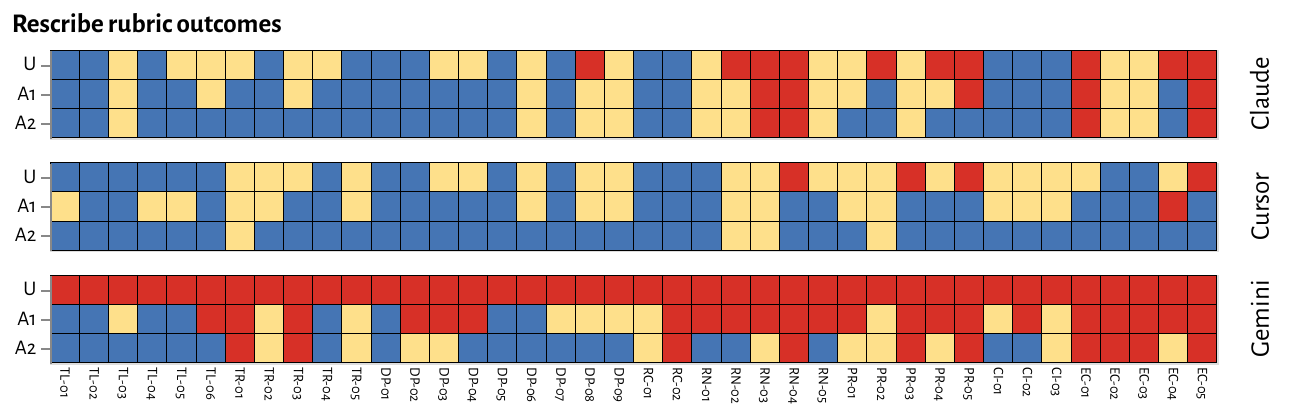}
    \caption{Revibeability performance for \textit{Rescribe}.}
    \label{fig:rescribe}
    \Description{The figure is a heatmap for the Rescribe revibe. Each row is the rubric’s results after each run (an aided and two unaided). Each column is a test. The cell colors red, yellow, and blue correspond to failure, partial success, and complete success, respectively. System rows are grouped to show changes in score over iterations. In this case, Claude goes from .55 to .7 to .78. Cursor goes from .65 to .78 to .95. Gemini goes from 0 to .33 to .61.}
\end{figure}

\begin{figure}[t]
    \centering
    \includegraphics[width=\linewidth]{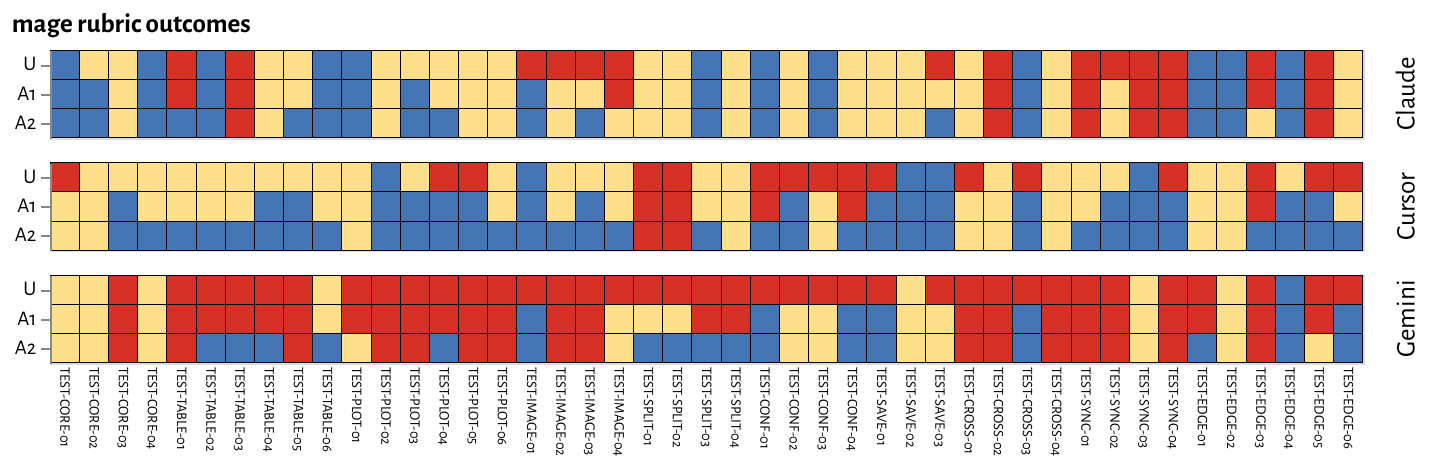}
    \caption{Revibeability performance for \textit{mage}.}
    \label{fig:mage}
    \Description{The figure is a heatmap for the mage revibe. Each row is the rubric’s results after each run (an aided and two unaided). Each column is a test. The cell colors red, yellow, and blue correspond to failure, partial success, and complete success, respectively. System rows are grouped to show changes in score over iterations. In this case, Claude goes from .48 to .57 to .66. Cursor goes from .54 to .77 to .84. Gemini goes from .1 to .3 to .51.}
\end{figure}

\begin{figure}[t]
    \centering
    \includegraphics[width=\linewidth]{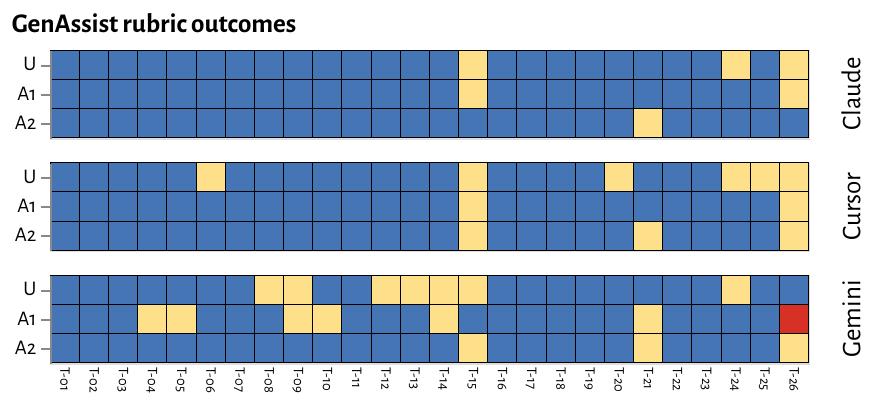}
    \caption{Revibeability performance for \textit{GenAssist}.}
    \label{fig:genassist}
    \Description{The figure is a heatmap for the GenAssist revibe. Each row is the rubric’s results after each run (an aided and two unaided). Each column is a test. The cell colors red, yellow, and blue correspond to failure, partial success, and complete success, respectively. System rows are grouped to show changes in score over iterations. In this case, Claude goes from .94 to .96 to .98. Cursor goes from .88 to .96 to .94. Gemini goes from .87 to .85 to .94.}
\end{figure}

Figures~\ref{fig:spellburst},~\ref{fig:rescribe},~\ref{fig:mage},~\ref{fig:genassist} reflect the disaggregated \rvb scores for each system we evaluated (the other results are provided in the Appendix). Each row represents a graded rubric (columns are individual tests). A red cell indicates a failure, yellow is partial success, and blue is full success. The results are grouped by agentic system, with the first row showing results for the unaided \rvb. The two remaining rows, A1 and A2, correspond to results after the first and second aided revibes.

\subsection{Qualitative Results \& Failure Analysis}
We reflect on a number of qualitative observations in revibing (graded \rvb rubrics are provided in the web supplement). 

\textbf{Near Implementations}---In some situations, it may be impossible to gain access to certain software or libraries for a perfect \rv. For example, \textit{Spellburst} used GPT 3.5 to generate p5 code (specifically, \texttt{gpt-3.5-turbo}, which first appeared in 2023). Because we provided the \rv projects with an Anthropic API key (and an indication that they could use an equivalent API), they generated code that called various Anthropic models. Depending on the goal for the \rv, this may not meet our needs. However, a more fundamental problem is that the previous APIs may no longer exist. As commercial models are retired, it may be impossible to perfectly replicate an implementation.

Additionally, some design elements in the original paper (and associated rubric tests) may focus on features that are less relevant for current models or libraries. For example, \textit{Spellburst} had error handling features (e.g., when generating non-functional p5 code). With improved models, such as the Claude Sonnet 3.5 (2024) or Sonnet 4 (2025), those failures do not appear in revibes. That is, the features designed to account for certain failures at the time the papers were created are no longer relevant. It is not clear that there is any strategy that would improve \rv performance in these cases--key software is simply unavailable.

In papers that used LLMs, we found that the agentic systems created their own prompts when none were documented. However, in some cases, even when the prompts were available (e.g., \textit{EvalGen}~\cite{validates}), the systems ignored them and wrote their own. This might indicate the need for a more explicit \rv prompt to force the agents to use existing material.

\textbf{Failed Features}---With \textit{mage}, we observed that none of the systems completely implemented the drag and drop functionality. Here, plot marks (e.g., bars) could be dragged to a `table' button to construct a filtered table in another notebook cell. This change would likely require modification of the underlying visualization library (the specific details were not obvious in the paper). The closest approach, that implemented by Cursor, allowed for using the native Vega-Lite interactions to select marks and a separate button (without dragging) to create a table. While the capabilities are analogous, the UX deviations might be limitations. Solving this may require prompting the systems to modify underlying libraries as needed. 

\textbf{Over-implementation}---When considering a paper, we found that some revibes over-implemented features. For example, in \textit{mage}, the system uses Vega-Lite as a visualization library and discusses the use of Altair (a different API) as future work. Claude actually implemented the Altair version. The \textit{mage} paper is also constructed in two phases. First, the authors built two interactive displays (one for tables and one for visualizations). After trying these, they asked participants to come up with additional ideas. The authors then implemented these as a proof of concept (e.g., image editing, train/test partitioning). These were not part of the original system. However, all three agentic systems implemented all the widgets. A more detailed prompt may be needed to focus the agentic system on the original system and not over-implement.

\textbf{Focus on Web Applications}---The \textit{mage} and \textit{B2} systems are interesting case studies in their target environment. Unlike many of our other revibes, they were not intended to be a web app. Rather, the systems were Jupyter plugins that provided custom widgets and tools that could be invoked for a range of tasks: to visualize the data (e.g., in tables, plots, or confusion matrices) or to export (e.g., images or CSV files). The authors of the \textit{mage} paper noted specifically that the familiarity of the actual Jupyter notebook was important to them: ``I think the fact that we had it in Jupyter, which is a familiar interface to data scientists, yeah, I think made them instantly see, okay, \ldots I already know all of this, and now he's showing me something new on top of that,'' In the case of mage, all the agentic systems built a new web app instead of a Jupyter plugin (notably, Cursor offered to build a notebook extension, but not as the default/recommended course of action). 

The author of B2 had a similar observation about the produced revibes. While largely successful in terms of features, the new versions would not provide the desired experience given the authors' goals. The author noted that they wanted to create a `polished' artifact that could be deployed within a familiar environment. Additionally, certain key features in B2 (e.g., reactive cells) worked \textit{because} the system was built on the specific Jupyter infrastructure. The \rv failed to implement this feature in a robust way by creating it from scratch. It remains an interesting question whether a \rv that does not meet the authors' design goals can nonetheless serve as a useful \rv for certain applications.

For mage, explicitly removing the prompt text ``Create this as a web app if you can, but you can use an alternative approach if the design calls for it (e.g., a Python backend).'' led Claude to focus on building a plugin. However, Cursor continued to prefer building its own Web app from scratch. This \textbf{implementation deviation} would need to be controlled for if the goal is closer adherence to the original design. For B2, we were able to partially \rv a Jupyter extension by modifying the prompt (see Figure~\ref{fig:b2extension}).

\textbf{Non-UI Features}---The \rvb rubric, as constructed, focuses on the interface and interactions. In some of our revibes, a main part of the contribution was in the backend or algorithms. For example, \textit{Rescribe} has a number of approaches to fit narrated descriptions in videos (e.g., through dynamic programming and optimizations). \textit{EvalGen}  also had algorithmic features (e.g., a selectivity-based grading sampler and alignment ranking mechanism). Current \rvb rubrics do not directly capture ways of testing these. Rather, they approximate this through interactive features. While the revibes may be ultimately successful, the feedback mechanism was not optimal. Rubrics that allow for other kinds of testing may reduce the number of feedback cycles needed.

In some cases, it may be impossible to exactly \rv a system without access to certain models or examples. In testing some of the systems (e.g., \textit{Rescribe}), we would have to find example files (e.g., videos). Without access to the exact (or `close enough') examples described in the paper, some systems were harder to test. In some cases, we observed the agentic tools create their own test data (e.g., audio files or images). In the case of \textit{Spellburst}, the system used fine-tuned examples (for a few shot prompting approach). This data was not available and certainly not revibeable as it used a crowdsourced collection mechanism. The author of the \textit{Spellburst} paper commented specifically on this as a potential issue (at the same time noting that, ``the design and interaction paradigm \ldots maps pretty much exactly to \ldots what we had.'').

One \rv we did not complete was for \textit{PromptPaint}~\cite{promptpaint}. While this system does meet our criteria (largely a software artifact), it utilized Stable Diffusion to produce generative images. While the UI worked, the agentic environment did not have access to the requisite GPUs. The tools would try to use external APIs (e.g., Gemini) as an alternative, but this led to a cascade of errors: recurring API model name changes, streaming incompatibilities with Flask/werkzeug, content policy blocks, and eventually an eventlet/kqueue conflict on macOS. The root cause was the mismatch between the paper's assumed infrastructure and our available resources. This suggests that revibeability has an implicit dependency on environment parity, and future work could explore how providing richer infrastructure (e.g., GPU access, containerized environments matching original development setups) might expand the scope of what vibe-coding can successfully reproduce.

\textbf{Rubric Observations}---One thing we noticed was that agentic systems would occasionally confuse `partial success' descriptions as `partial failures.' The difference for a human reader may be clear if they read what defines these criteria in the rubric. For example, it may not be obvious to the agent if `code prepended' should be considered a failure or success if it does not observe the full definition. In fact, we saw that some agents would use the \textit{grep} tool to search for `Observed results' lines in rubric files. Doing so would mean they were not considering the actual test definition lines. In this case, the failure might not be corrected until the tester provides additional clarification (e.g., `code prepended incorrectly'). Additional prompt instructions (e.g., `consider definitions of success or failure in the rubric') might lead to better results.

A related rubric issue is with ambiguously described features. In some cases, the specific implementation of certain visual features was not clear. This may be due to space constraints, author oversight, or simply that it was not important. Screenshots of systems, in particular, were often cropped or too small for both the agentic system and tester/developer to read. In these instances, it was difficult to know how the UX should look. Clearly, this is not simply a problem for revibes. However, a better rubric-construction prompt may indicate areas of ambiguity.

\textbf{Author Memory}---One unanticipated observation from our interviews was that authors could not always remember how they implemented a feature or what was in the paper. For example, with \textit{mage}, the details on how certain variables were retained or how drag-and-drop was implemented were not easily recalled. With \textit{Spellburst}, a feature built internally--and which the author recalled as being important--was not actually described in the paper. This points to the possible benefit of creating a \rv or \rvb rubric when writing the paper to ensure proper alignment. 

\subsection{Rubric Evaluation}\label{sec:rubrics}
\begin{figure}
    \centering
    \includegraphics[width=\linewidth]{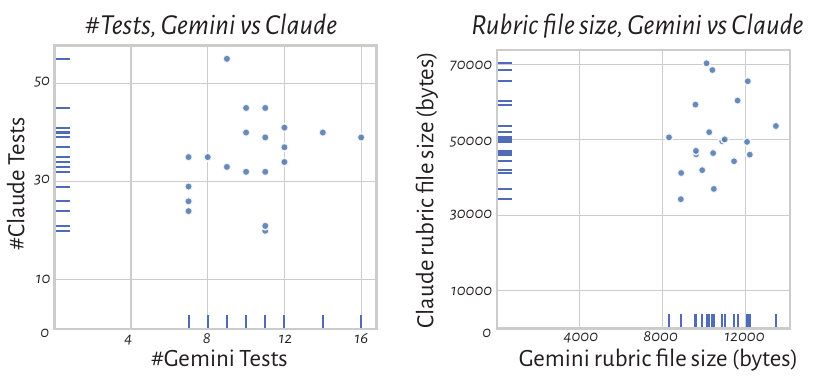}
    \caption{A comparison of \rvb rubric test counts and file size. Each dot represents a system.}
    \label{fig:rubric}
    \Description{The figure shows two side-by-side scatterplots. The scatterplot on the left contrasts the number of tests generated by Gemini versus those generated by Claude for the same system (20 points total). There is a correlation between the two, but Gemini generates 7 to 16 tests, whereas Claude generates 2 to 3 times as many for each system. The second scatterplot contrasts the file sizes of the rubrics. Again, we see 2 to 3 times larger rubrics for Claude over Gemini.}
\end{figure}

In the experiments described above, we opted to use Claude Code as the generator for \rvb rubrics. In pilot testing, we found that Claude Code produced the most detailed rubrics. To confirm this observation, we took the original 20 systems (see Appendix~\ref{sec:systems}) collected as potential \rv targets and applied the rubric construction prompt through both Claude Code and Gemini. We measured the number of tests in each rubric, the number of steps in each test, and the length of the rubric (as a proxy for detail). Figure~\ref{fig:rubric} plots the number of tests per rubric (left) and file length (right). 

Rubric lengths are weakly correlated (.26 and .25 through Pearson correlations for test count and file sizes, respectively). However, Claude code produces rubrics with 3.5 times as many tests as Gemini (mean of 35.1 tests vs 10.25 tests). Similarly, Claude Code produces rubrics nearly 5 times longer than Gemini (mean of 50694 bytes versus 10570). Finally, Claude Code tests are on average 1.6 times longer than Gemini (4.7 steps versus 2.8 steps). While correlated, test counts, file sizes, and the number of steps in Gemini and Claude-produced rubrics are significantly different ($p < 0.001$ based on a paired t-test).

It is difficult to conclusively determine from the sizes of the rubrics whether one is better than the other. It is possible that there is a `core' set of tests that sufficiently tests core functionality \textit{and} provides enough information to the coding environment to make corrections. The original authors or \rv coders may not be able to identify the optimal subset that satisfies these two constraints. Automated optimization through ablation testing may be a possible approach. While we leave the optimization problem to future work, a simpler bootstrap analysis through resampling (see Appendix~\ref{sec:analysis}) may give better bounds on \rvb scores. The generated intervals provide a different collection of tests. It is also possible that additional agentic prompts may be useful in finding overlaps and recommending deletion. For now, over-testing may be reasonable.

One possible complication in our approach is that it might introduce a confound: Claude Code may identify the same set of interactive features in the paper when creating the rubric as when it produces the \rv. This may lead to it performing better because it also generated the test. However, we note that the prompts for rubric creation are very different from code construction. Additionally, we found that Cursor produced good revibes with the Claude-crafted rubrics. Given the sophistication and detail in the Claude rubrics, we opted to use them for our study.

%% file: 07_discussion.tex
\section{Discussion}~\label{sec:discussion}

\subsection{Limitations and Future Work}
The work described here is a floor, not a ceiling, for what can be achieved through agentic coding. First, it is also an open question of what fraction of historical UIST papers are revibeable. The research team assessed the roughly 210 papers from UIST 2025 and found that 65\% were primarily software papers that used off-the-shelf hardware and had at least the potential to be revibed. Of the remaining papers, 15\% were partially-revibeable. They had both a significant hardware (e.g., novel sensor) and software contribution (e.g., an inference method). The remaining papers were entirely hardware contributions and were likely not revibeable. We can expect that agentic programming environments will only improve over time. The author of B2, who now develops agentic tools, was convinced that most of the missing features in the \rv could be completed with additional prompts. Additionally, in using CLI-based agentic environments, we are already limiting the performance of the tools. Various environments offer automated browsing features that might drive testing and self-correction (e.g., Google Antigravity~\cite{antigravitybrowser} or Cursor~\cite{cursorbrowser}). Other approaches to visual GUI testing (e.g.,~\cite{qassem25}) may also support automated \rvb testing--allowing for fully unaided revibes. Better MLLMs/VLMs that can process supplementary videos or images may further improve revibes in the future (e.g., some design decisions are only documented in images). 

Second, our focus in this work has been on vibe-coding. The developer/tester does not ever touch the code in these reimplementations. In actual use, we might expect that researchers will try to complete the reimplementation if the \rv is insufficient. In these cases, we would expect the researcher to guide the coding process towards implementations that match their scientific intent. Naturally, this raises possible risks if researchers are not careful in how they use the approach. Strong scientific integrity and good experimental design will be critical. What are the tradeoffs for incorporating more researcher efforts, and how to best utilize researcher inputs are open questions for future research.

Third, the prompts we describe in this paper were developed through piloting. They were designed to be simple on purpose so that we could understand how the agentic environment's own planning and implementation prompts would work. Future work may try to identify better prompts for \rv, particularly those that work well with specific agentic environments or artifact types. As discussed in the qualitative findings, the prompts we use are effective at producing web applications. We expect that older UIST papers, which are not web-based or have a hardware component, may require tweaks to the prompts. 

Finally, while we have found that our prompts for rubric construction produce satisfactory results, they can certainly be improved through a combination of author feedback and better automation. When studying the rubric, the author of the B2 system could readily identify those tests that connected to critical system features. This suggests that author involvement is highly advantageous. Targeting key features is one improvement, but expanding the rubrics beyond web interactions may also be critical. In some cases, the core contribution of a paper may be in a back-end algorithm and not in the UI. Producing \rvb rubrics that can test those features directly may be necessary. Additionally, there are many ways to create system rubrics (e.g., perceptual, lexical, operational, and compositional~\cite{majrashi2023inter}). Better rubric generator prompts may create tests that target different system aspects.

\subsection{Applications of Revibeability}
While we have used \rvb in a very specific way in this paper, we note that there are many possible future applications for the metric. For example:

\begin{itemize}[noitemsep,leftmargin=*]
    \item We can compare \rvb scores to determine which agentic system or model works better for revibes (in general or for a specific system). This in itself can become a benchmark task to incentivize future models to improve their revibing capabilities.
    
    \item We can track improvements in \rvb over time as agentic systems change by using the same rubric (i.e., using specific systems as a benchmark).
    
    \item We can provide a \rvb score for any given paper (e.g., by averaging across multiple agentic programming tools), providing a new kind of badge or serving as a reviewing approach\footnote{Though we emphasize that a low or high score does \textit{not} indicate anything about the technical sophistication of the work, just how easily we can \rv it.}.
    
    \item We can test different variants of the same paper and rank them based on \rvb. This would enable authors to check which version of their paper is sufficiently detailed.
    
    \item A conference can identify and recommend optimal paper features (e.g., ideal length, required sections) based on which features correlate with high \rvb.
\end{itemize}

\subsection{Changing Technical HCI}
Within academia, we often espouse the values of open science~\cite{cohoon2021norms}, but fail to provide the software needed to make that happen. Sharing software is neither required nor rewarded. In fact, it sometimes creates additional work. Academics who want to deploy code have to worry about maintenance, documentation, and IP issues (see~\cite{Echtler18}). In some ways, revibing fixes this. As long as an author checks their paper for \rvb, or provides a \rvb rubric, it becomes far easier for others to build upon. Additionally, we have a tendency to forget prior work in HCI~\cite{oppenlaender2025keeping}. Part of forgetting may be due to artifacts becoming harder to find and run. Revibing can make systems evergreen and can work not just on papers but existing code bases. The result is that existing artifacts (revibed or not) can be updated to run in modern environments.

Another potential future implication of revibing is that authors may be encouraged to provide sufficient documentation or rubrics for others to reimplement their system. We can find some inspiration from the open source community. For example, \textit{whenwords}~\cite{whenwords}, a library for `relative time formatting,' offers no code. Instead, it provides a human-readable specification and a set of tests. The intent is for developers to provide both of these to an agentic programming environment, and the output would be a complex, functional library in the developer's language of choice. The authors of the library do not offer any particular implementation (though they do note that they have tested their approach on a few languages). Similarly, authors of a paper can provide additional descriptive details of their system (or screenshots) or \rvb rubrics (e.g., in supplementary materials). These can be generated automatically with minimal effort from the authors, just as we use a prompt for automatic rubric generation. 

Within the UIST community, we have developed a sense of what constitutes a technical contribution and how it should be evaluated~\cite{fogarty2017code,greenberg08,ledo18,olsen07}. Revibing, as we demonstrate in the paper, may challenge these existing norms and practices. For example, it may become more difficult to justify why one does not do more comparisons with strong baselines when evaluating their work. However, as the possibility for more A-B evaluations increases, so do the possible problems in their implementation~\cite{dix2010human}. A strong baseline does not automatically mean it is a valid `B' without additional controls. Furthermore,  Alan Dix has argued~\cite{dix2010human} for specialization in the HCI community: people who are good at building versus people who are good at experimenting. Until recently, not being able to build meant that it was difficult to run certain kinds of experiments. The feasibility of revibing may break this division and open doors to new kinds of work within the HCI community. For example, people who are good at experiments but not good at building \textit{can} run new kinds of evaluations by reimplementing systems. Revibing approaches may require additional training and the development of community norms. A poorly constructed \rv may appear to outperform a new system, but this may be misleading. Authors would need to carefully argue that the reimplementation is appropriate. Simply reporting the \rvb score to demonstrate that some implementation is a strong baseline would not be sufficient. In some ways, this is not a new problem. One would need to do this whether the reimplementation was agentic or not. However, there are increased risks if reimplementations through \rv become the norm rather than the exception.

We must also consider the possible negative impact that revibing may have on HCI research practices. It is possible that if the community is more inclined to use strong baselines, there may be natural incentives to create work that is revibeable (so that it is built upon and cited). This may be undesirable from the community perspective. For example, reviewers might argue to reject a paper because was is not revibeable in some naive way. There are many types of contributions a technical HCI artifact can make and not all are represented in the artifact itself. Just as naive abuse of \rv is problematic for authors, it may similarly be problematic for reviewers. Developing community norms and training for reviewers will also be needed if revibing is normalized.  It is also possible that authors may be further disincentivized from releasing their actual artifacts. It is worth noting that our approach for evaluating \rvb focuses only on ``sufficiently covering intended functionalities.'' and does not provide a full picture of the gaps between revibed artifacts and the original ones. If any of these gaps are systematic and unattended, they could bias scientific advancement if revibed artifacts are to become an integral part of HCI evaluation.  Constant revibing may also raise broader concerns, such as the resource and environmental costs of agentic programming. 

Many papers in technical HCI are \textit{greenfield programming}~\cite{hopkins2008eating}. That is, they are constructed largely from scratch (modulo libraries and toolkits) and are not part of large codebases (brownfield programming). There are various systemic and practical reasons why this is the case. We reward novelty and simple proofs-of-concept over hard-coded deployments. It is hard to anticipate the impact of agentic development on this practice going forward. On the one hand, we see an explosion of new contributions because it is far simpler to develop technical artifacts. The result may very well be that we continue to see a vast majority of papers building through vibe-coded greenfield programming. On the other hand, the commercial recognition of the importance of brownfield programming may lead to better agentic tools to modify existing, large-scale codebases. It is possible that this will lead to other ways of producing UIST contributions. Any use of \rvg will need to adapt to the kinds of systems we produce.

One final consideration for the technical HCI community is how \rvg may alter the way we train our students. We already use reimplementation as a learning tool~\cite{belen2025seems,syeda2024vis}. Revibing has clear implications for how to structure these kinds of assignments (will reimplementation actually be challenging?). However, it also allows us to develop new kinds of educational materials. For example, students can be tasked with not just reading a paper but \rvg and extending it.

%% file: 08_conclusions.tex
\section{Conclusions}
In this work, we focus on the possibility of reimplementing technical HCI artifacts through agentic coding. Specifically, we introduce the concept of a \rv (to recreate the system without coding from the paper) and \rvb (a measure of revibing success). We recreate 10 different systems from recent UISTs using three different agentic coding environments. The results are promising. Revibed systems are good reimplementations of many of the original system's interface features (and in some cases, perfect). This raises the possibility of using revibed artifacts in evaluation. The possibility of rapid reimplementation opens up other avenues for education and knowledge production in the technical HCI field.

%% file: 98_ack.tex
\begin{acks}
We would like to thank Paul Resnick and Vidya Setlur for feedback on this project. We are grateful to the authors who participated in our interviews. Finally, we thank the anonymous reviewers for their valuable feedback and suggestions.
\end{acks}

%% file: 99_appendix.tex
\section{Appendix}

\subsection{Systems}\label{sec:systems}

For our rubric experiments, the following papers were used: \textit{B2}~\cite{b2}, \textit{Cells, Generators, and Lenses}~\cite{cellsgens}, \textit{CoLadder}~\cite{coladder}, \textit{EvalGen}~\cite{validates}, \textit{GenAssist}~\cite{genassist}, \textit{mage}~\cite{mage}, \textit{Memory Sandbox}~\cite{memorysandbox}, \textit{Patchview}~\cite{patchview}, \textit{ProactiveVa}~\cite{proactiveva}, \textit{Rescribe}~\cite{rescribe}, \textit{Sketch-n-Sketch}~\cite{sketchnsketch},  \textit{Spellburst}~\cite{spellburst}, \textit{Synergi}~\cite{synergi}, \textit{Vizability}~\cite{vizability}, \textit{WaitGPT}~\cite{waitgpt}, \textit{WorldScribe}~\cite{worldscribe}, \textit{WorldSmith}~\cite{worldsmith}, and \textit{XCreation}~\cite{xcreation}.

\begin{figure}[h]
    \centering
    \includegraphics[width=\linewidth]{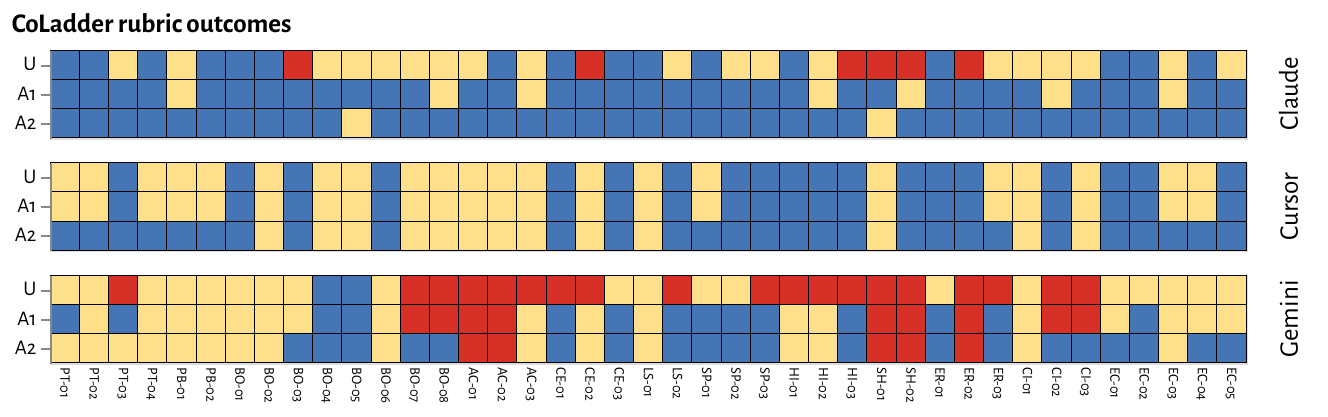}
    \caption{Revibeability performance for \textit{CoLadder}.}
    \label{fig:coladder}
        \Description{The figure is a heatmap for the CoLadder revibe. Each row is the rubric’s results after each run (an aided and two unaided). Each column is a test. The cell colors red, yellow, and blue correspond to failure, partial success, and complete success, respectively. System rows are grouped to show changes in score over iterations. In this case, Claude goes from .62 to .91 to .98. Cursor goes from .73 to .73 to .84. Gemini goes from .29 to .56 to .68.}
\end{figure}

\begin{figure}[h]
    \centering
    \includegraphics[width=\linewidth]{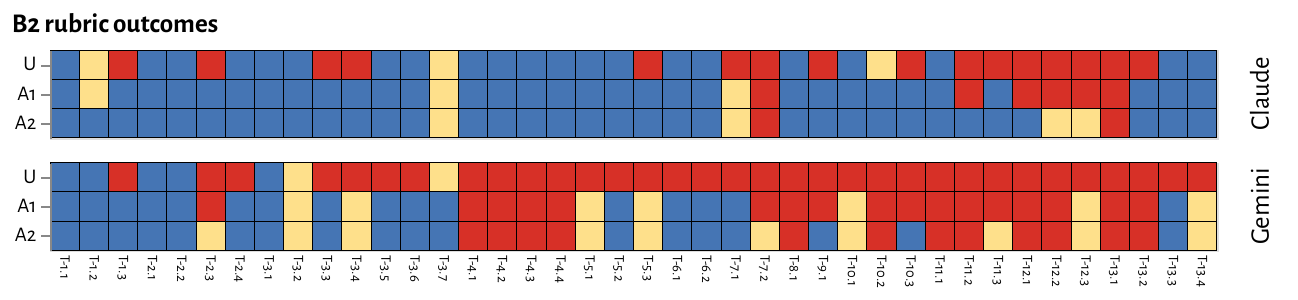}
    \caption{Revibeability performance for \textit{B2}}
    \label{fig:b2}
\Description{The figure is a heatmap for the B2 revibe. Each row is the rubric’s results after each run (an aided and two unaided). Each column is a test. The cell colors red, yellow, and blue correspond to failure, partial success, and complete success, respectively. System rows are grouped to show changes in score over iterations. In this case, Claude goes from .56 to .81 to .9. Gemini goes from .15 to .59 to .58.}
\end{figure}

\begin{figure}[h]
    \centering
    \includegraphics[width=\linewidth]{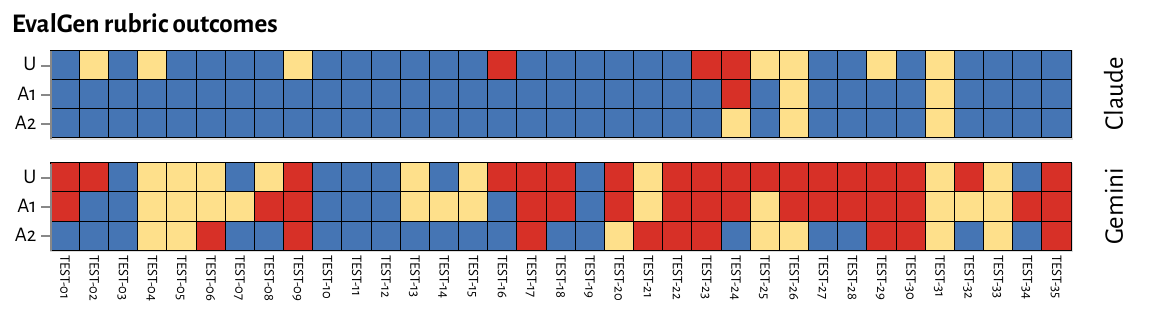}
    \caption{Revibeability performance for \textit{EvalGen}}
    \label{fig:validates}
    \Description{The figure is a heatmap for the EvalGen revibe. Each row is the rubric’s results after each run (an aided and two unaided). Each column is a test. The cell colors red, yellow, and blue correspond to failure, partial success, and complete success, respectively. System rows are grouped to show changes in score over iterations. In this case, Claude goes from .81 to .94 to .96. Gemini goes from .36 to .37 to .64.}
\end{figure}

\begin{figure}[h]
    \centering
    \includegraphics[width=\linewidth]{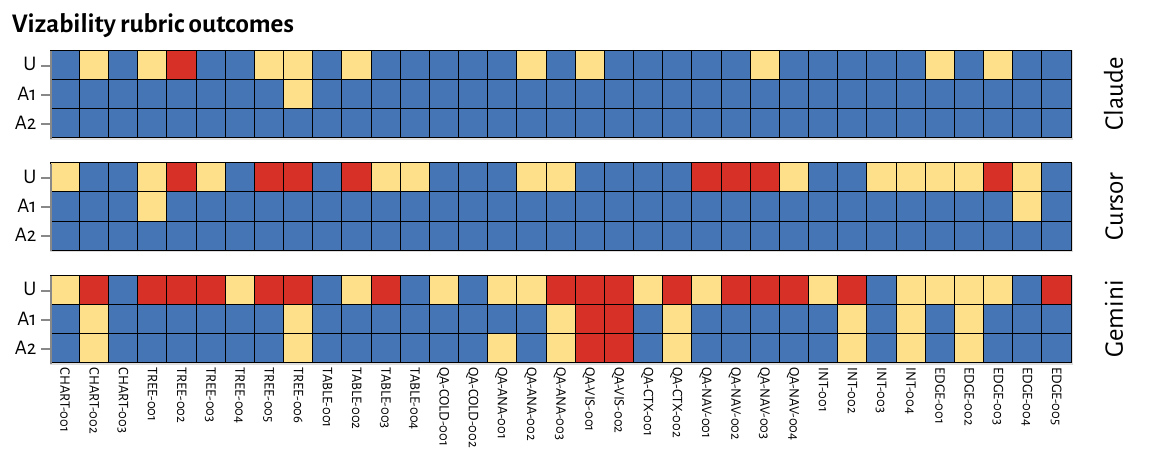}
    \caption{Revibeability performance \textit{Vizability}.}
    \label{fig:vizability}
    \Description{The figure is a heatmap for the Vizability revibe. Each row is the rubric’s results after each run (an aided and two unaided). Each column is a test. The cell colors red, yellow, and blue correspond to failure, partial success, and complete success, respectively. System rows are grouped to show changes in score over iterations. In this case, Claude goes from .94 to .96 to .98. Cursor goes from .88 to .96 to .94. Gemini goes from .87 to .85 to .94.}
\end{figure}

\begin{figure}[h]
    \centering
    \includegraphics[width=\linewidth]{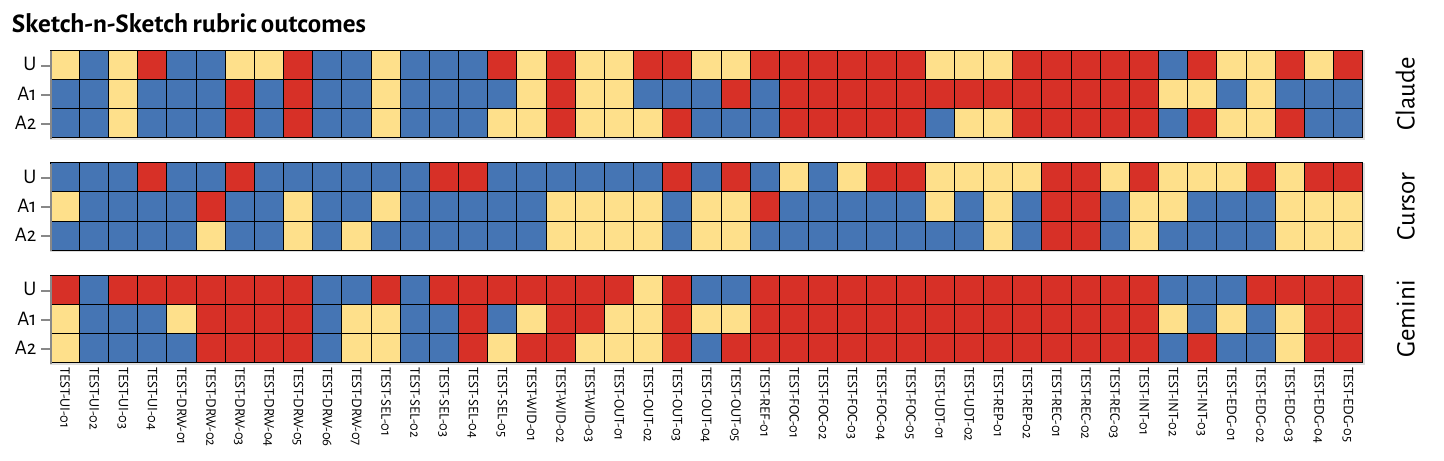}
    \caption{Revibeability performance for \textit{Sketch-n-Sketch}.}
    \label{fig:sketchnsketch}
    \Description{The figure is a heatmap for the Sketch-n-Sketch revibe. Each row is the rubric’s results after each run (an aided and two unaided). Each column is a test. The cell colors red, yellow, and blue correspond to failure, partial success, and complete success, respectively. System rows are grouped to show changes in score over iterations. In this case, Claude goes from .83 to .99 to 1. Cursor goes from .58 to .97 to 1. Gemini goes from .36 to .84 to .83.}
\end{figure}

\begin{figure}[h]
    \centering
    \includegraphics[width=\linewidth]{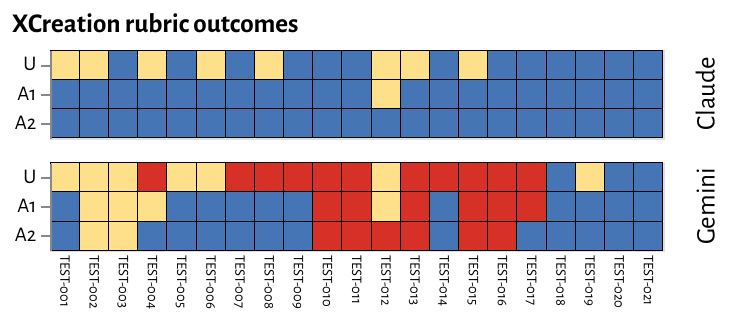}
    \caption{Revibeability performance for \textit{XCreation}}
    \label{fig:xcreation}
    \Description{The figure is a heatmap for the XCreation revibe. Each row is the rubric’s results after each run (an aided and two unaided). Each column is a test. The cell colors red, yellow, and blue correspond to failure, partial success, and complete success, respectively. System rows are grouped to show changes in score over iterations. In this case, Claude goes from .81 to .98 to 1. Gemini goes from .31 to .62 to .67.}
\end{figure}

\begin{figure}[h]
    \centering
    \includegraphics[width=\linewidth]{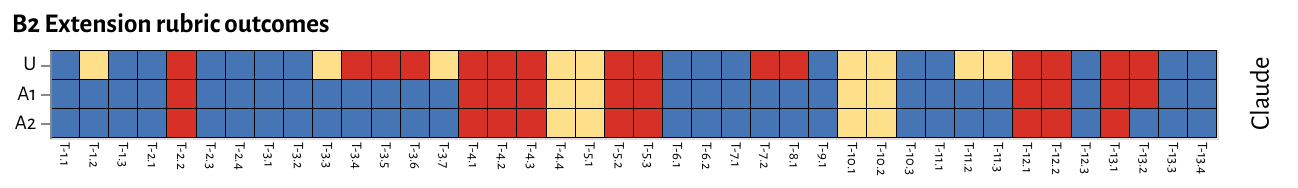}
    \caption{Revibeability performance for \textit{B2} (Jupyter Extension version)}
    \label{fig:b2extension}
    \Description{The figure is a heatmap for the B2 Jupyter extension revibe. Each row is the rubric’s results after each run (an aided and two unaided). Each column is a test. The cell colors red, yellow, and blue correspond to failure, partial success, and complete success, respectively. System rows are grouped to show changes in score over iterations. In this case, Claude goes from .51 to .7 to .73.}
\end{figure}

\subsection{Model vs Agent}\label{sec:cursor}
\begin{figure}[h]
    \centering
    \includegraphics[width=\linewidth]{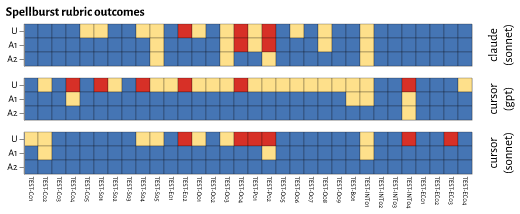}
    \caption{Comparative revibeability performance for \textit{Spellburst} (agent $\times$ model)}
    \label{fig:spellburstspecial}
    \Description{The figure is a heatmap for the Spellburst revibe. It summarizes the contrast between two models (GPT 5.4 xhigh and Claude Sonnet 4.6) and two agents (Cursor and Claude). Each row is the rubric’s results after each run (an aided and two unaided). Each column is a test. The cell colors red, yellow, and blue correspond to failure, partial success, and complete success, respectively. System rows are grouped to show changes in score over iterations.}
\end{figure}

\begin{figure}[h]
    \centering
    \includegraphics[width=\linewidth]{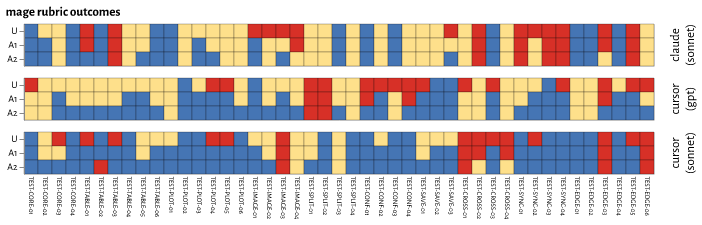}
    \caption{Comparative revibeability performance for \textit{mage} (agent $\times$ model)}
    \label{fig:magespecial}
        \Description{The figure is a heatmap for the mage revibe. It summarizes the contrast between two models (GPT 5.4 xhigh and Claude Sonnet 4.6) and two agents (Cursor and Claude). Each row is the rubric’s results after each run (an aided and two unaided). Each column is a test. The cell colors red, yellow, and blue correspond to failure, partial success, and complete success, respectively. System rows are grouped to show changes in score over iterations.}
\end{figure}

It is challenging to definitively conclude that one agentic architecture is better than another. Because we do not have access to the system/agent prompts for closed-source systems, the specific impacts of these prompts are hard to disentangle. In Figures~\ref{fig:spellburstspecial} and~\ref{fig:magespecial}, we reiterate the results from Claude (with \texttt{claude-sonnet-4-6}) and Cursor (\texttt{gpt-5.4-xhigh}) but add Cursor with \texttt{claude-sonnet-4-6} to those.

Through this experiment, we cannot conclusively say that one system is a better `user' of a particular model or which model is best to use given the same agentic prompts. However, the results do indicate that the internal prompts of the agentic systems can have an impact on \rv performance, and additional research is warranted.

\subsection{Rubric Example}\label{sec:rubric_example}

Example `tests' produced for the Spellburst system~\cite{spellburst}. These were produced using Claude Code (\texttt{Sonnet 4.6}). The first represents a simpler test case. The second, which appears towards the end of the rubric, is more complex.

\subsubsection{TEST-S05}

\begin{tcolorbox}[
  colback=notebg,     
  colframe=notebg,    
  boxrule=0pt, arc=0pt,
  left=2pt, right=2pt, top=2pt, bottom=2pt,
  breakable]
\ttfamily\small
\begin{Verbatim}[fontsize=\small, breaklines=true, breakanywhere=true, breaksymbolleft={}]
### TEST-S05 - Node Deletion

**ID:** TEST-S05
**Prerequisites:** At least a root node and one child sketch node exist on the canvas.

**Steps:**
1. Identify a child sketch node that has a parent node and at least one descendant node.
2. Click the `×` close button in the node's header.
3. Observe: (a) whether the node is removed, (b) what happens to its descendants, and (c) what happens to the edge connecting parent to descendants.

**Expected Result:**
The deleted node is removed from the canvas. Its descendants are not deleted - they remain. The descendants are reattached to the deleted node's parent (so the edge connects the grandparent directly to the grandchildren). The overall graph remains navigable.

**Success Criteria:**
- **Full Success:** Node disappears; descendants remain and are reattached to the grandparent, preserving the exploration history as described in Section 5.2.3.
- **Partial Success:** Node is deleted but descendants are also deleted (cascading delete), or descendants are orphaned (disconnected) rather than reattached.
- **Failure:** The x button is absent, clicking it has no effect, or deletion crashes the application.

**Observed result:**
\end{Verbatim}
\end{tcolorbox}

\subsubsection{TEST-INT01}

\begin{tcolorbox}[
  colback=notebg,     
  colframe=notebg,    
  boxrule=0pt, arc=0pt,
  left=2pt, right=2pt, top=2pt, bottom=2pt,
  breakable]
\ttfamily\small
\begin{Verbatim}[fontsize=\small, breaklines=true, breakanywhere=true, breaksymbolleft={}]
### TEST-INT01 - Full Exploratory Session (Ash Scenario)

**ID:** TEST-INT01
**Prerequisites:** Fresh canvas; LLM backend reachable. Allow approximately 15-20 minutes for this test.

**Steps:**
1. Create a root sketch node via `+ New Sketch`.
2. Open the code editor and paste in a simple p5.js concentric circles sketch.
3. Verify the output renders (concentric circles visible).
4. From the root node, use Duplicate to create a child; open its code editor and change a numeric parameter (e.g., number of circles). Verify the change renders instantly.
5. From the root node, create a second branch using Duplicate; change a color parameter in this branch. Verify it shows a different color from Step 4's branch.
6. From the root node, create a Modify operator with the prompt: *"Pretend the circles are the orbit of planets of our solar system. Add some fake planets and a sun."* Click Generate. Verify a new animated sketch with planets is created.
7. On the generated planet sketch, type a partial prompt in a new Modify node and use autocomplete to select a suggestion. Generate. Verify a new child sketch is created.
8. On a subsequent sketch in the topographic/noise branch, use the semantic parameter sliders to adjust parameters. Verify real-time output changes.
9. Select two sketches from different branches and Merge them. Verify a new combined sketch is generated with a merge comment in its code.
10. Use the Extract operator on one sketch to isolate a visual property.
11. Review the overall canvas: verify the auto-layout shows a hierarchical tree from left to right matching the structure in Figure 2.

**Expected Result:**
By the end of the session, the canvas shows a tree of interconnected sketch and operator nodes. The session mirrors the workflow described in Section 4, producing a range of visually distinct sketches (concentric circles -> planets -> topographic maps -> color gradients -> merged result). All operators functioned correctly during the session.

**Success Criteria:**
- **Full Success:** All 11 steps complete without error. The resulting canvas resembles Figure 2 in structure (hierarchical, left-to-right). At least 8 distinct sketch nodes are visible. All operators (Modify, Duplicate, Merge, Extract) produced correct results.
- **Partial Success:** 7-10 steps complete correctly; 1-2 operators fail or produce unexpected results; canvas structure is approximately tree-like but with some layout issues.
- **Failure:** Fewer than 6 steps complete; multiple operators fail; the canvas layout is chaotic; or the session is interrupted by application errors.

**Observed result:**
\end{Verbatim}
\end{tcolorbox}

\subsection{Rubric Analysis}\label{sec:analysis}

\begin{figure}
    \centering
    \includegraphics[width=.9\linewidth]{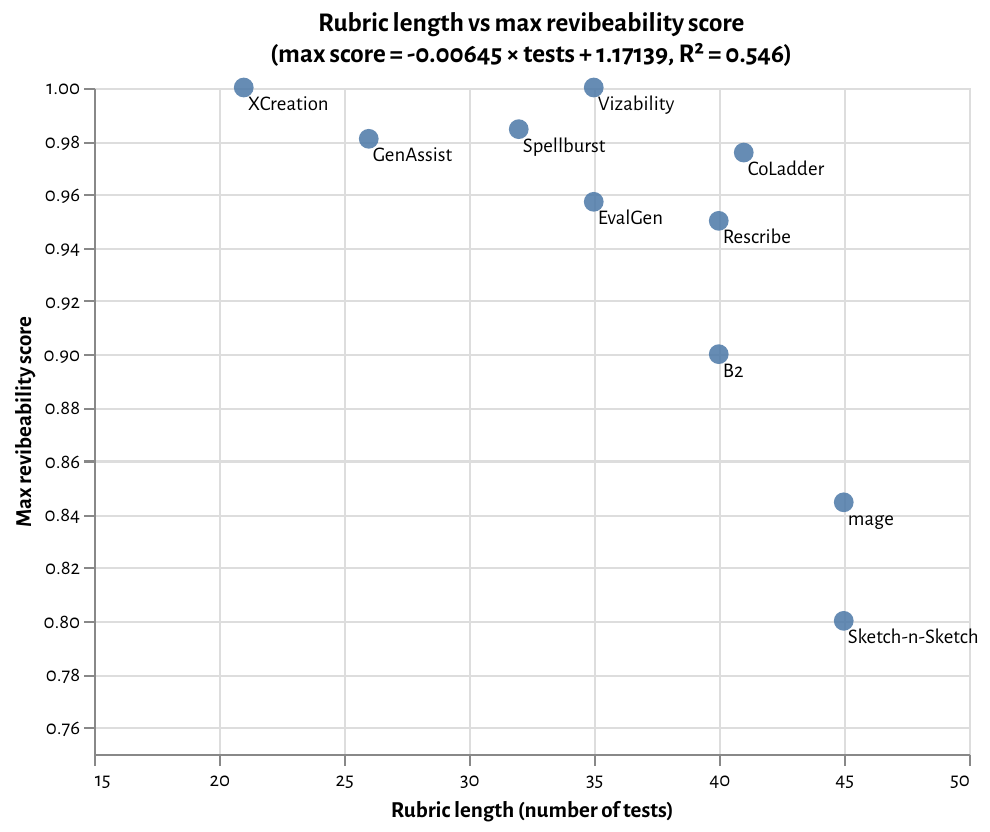}
    \caption{Correlation between best revibeability score and the length of the rubric.}
    \label{fig:correlation}
    \Description{The figure is a scatterplot showing a rough correlation between rubric length and top revibeability scores. Each point is a system. The plots shows a negative correlation between the two. As test length increases, revibeability score decreases.}
\end{figure}

Figure~\ref{fig:correlation} illustrates the correlation between \rvb rubric length and max scores achieved. The data is correlated (Pearson $r=-.74$, $p<.05$).

\begin{figure}[h]
    \centering
    \includegraphics[width=.9\linewidth]{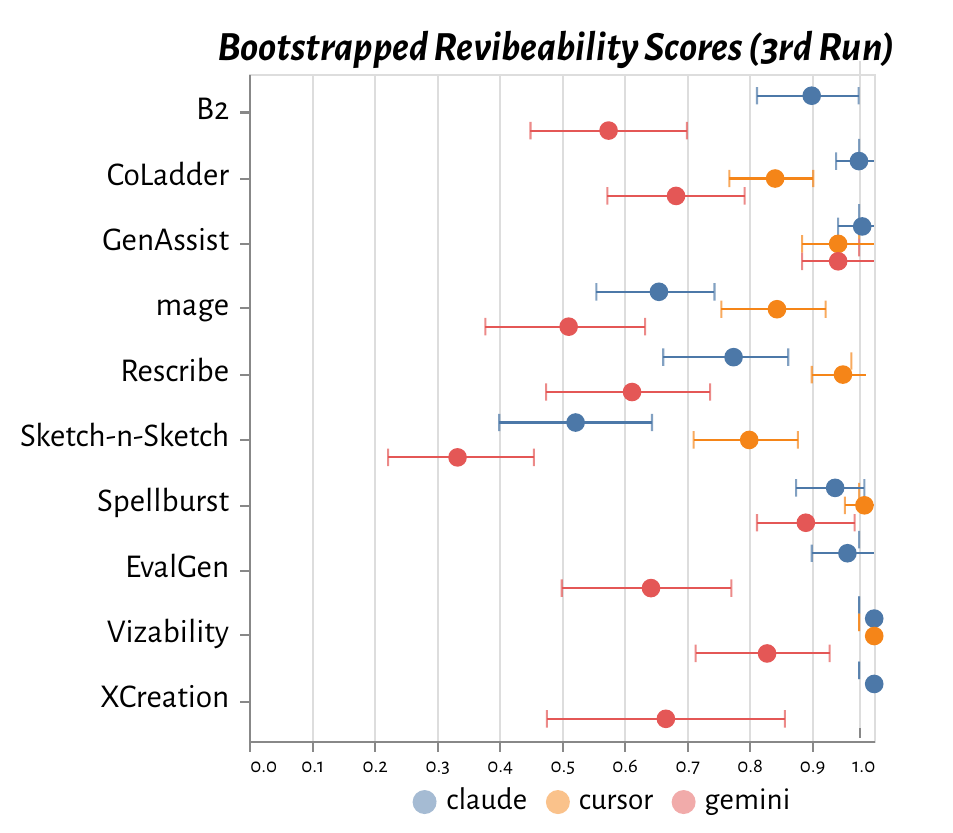}
    \caption{95\% CI calculated on the last run for each system by bootstrapping (1000 resamples)}
    \label{fig:bootstrap}
    \Description{The figure is a plot showing the 95\% confidence intervals for each \rv in the original paper. This is done through a bootstrap sampling (1000 times). The plot shows that the scores vary slightly around for \rv but the scores are roughly aligned with the main results.}
\end{figure}

To test the sensitivity of the results to the \rvb tests generated by the system, we perform a bootstrap analysis (see Figure~\ref{fig:bootstrap}). This was determined by randomly resampling (with replacement) tests from the 3rd run (the last aided revibe).

\subsection{Example Implementation Reports}

The following is Claude's report after the initial (unaided) \rv prompt for the \textit{Rescribe} system~\cite{rescribe} (Sonnet 4.6): 

\begin{tcolorbox}[
  colback=notebg,     
  colframe=notebg,    
  boxrule=0pt, arc=0pt,
  left=0pt, right=0pt, top=0pt, bottom=0pt,
  breakable]
\ttfamily\small
\begin{Verbatim}[fontsize=\small, breaklines=true, breakanywhere=true, breaksymbolleft={}]
Rescribe - Implementation Complete

The full system is implemented as a web app:

What was built

Backend (FastAPI + Python)
- audio_analysis.py - faster-whisper transcription + librosa gap detection (silence/music/extendable)
- nlp_candidates.py - spaCy-based candidate description generation (drops adjectives, prepositional phrases, compounds
while protecting film phrases and quoted text)
- scoring.py - Claude claude-haiku-4-5-20251001 coherence scoring + informativeness (noun TF-IDF) + edit quality (cut count)
- optimizer.py - greedy gap-fitting optimizer for inline/extended/extended-inline modes
- renderer.py - pydub audio mixing with edge-tts -> gTTS fallback for TTS
- main.py - 11 REST endpoints for the full workflow

Frontend (React + TypeScript + Vite)
- Upload screen - drag-and-drop video upload
- Video panel - player + custom timeline showing speech, gaps (colored by type), draft and rendered descriptions
- Transcript pane - time-aligned words + clickable gap blocks; descriptions shown inline in grey
- Description pane - cards with thumbnails, text editing, microphone recording, lock controls (text/time/presence), NLP shortening suggestions, and post-render word-level toggling + short/long slider
\end{Verbatim}
\end{tcolorbox}

Revision post-`feature check' prompt:

\begin{tcolorbox}[
  colback=notebg,     
  colframe=notebg,    
  boxrule=0pt, arc=0pt,
  left=0pt, right=0pt, top=0pt, bottom=0pt,
  breakable]
\ttfamily\small
\begin{Verbatim}[fontsize=\small, breaklines=true, breakanywhere=true, breaksymbolleft={}]
Everything builds cleanly. Here's a summary of all the missing features that were added:
---
Features Added

Backend algorithmic improvements

audio_analysis.py
- Beat detection using librosa.beat.tempo - music gaps are now classified as extendable_music only if tempo >= 60 bpm AND duration >= 10s (paper: >=60 bpm, >=30s), rather than a simple duration check
- Added transcribe_audio() for transcribing recorded description audio clips to get word-level timestamps

nlp_candidates.py
- Video-specific phrase protection - scans source transcript + all descriptions for 2/3/4-grams occurring 3+ times with at least one verb/noun; these are kept together and cannot be split
- On-screen text prefix detection - text following "text:", "title:", "credit:", "text on screen:", "title card:" prefixes is fully protected from any shortening
- Quotation-mark locking - fully quoted text is detected and returned as a single verbatim candidate

optimizer.py (complete rewrite)
- DP-based placement with proper cost function: C(di) + P(di,t,l) per the paper
- Placement penalties: speech overlap = $\infty$, previously placed description overlap = $\infty$, within 0.3s of boundary = +10,
extension per second (extended-inline) = +1
- Candidate selection - optimizer now receives scored candidates and picks the best (text, placement) pair globally
- Locked constraint enforcement - locked_text, locked_time, locked_presence all respected

renderer.py
- Word-level audio editing (edit_audio_by_word_selection) - cuts recorded audio at word boundaries and concatenates kept segments with short crossfades, so user word selections don't require re-recording
- Improved ambient noise extension using random segment sampling (avoids rhythmic artifacts)

main.py
- Candidate scoring integrated into render - every render call now generates NLP candidates and scores them with Claude before optimization
- Word selection endpoint (POST /descriptions/{id}/word-selection) - saves included words and returns re-cut audio
- Batch export endpoint (POST /export-all) - renders inline, extended, and extended-inline from one set of descriptions
- Recording transcription - uploaded recordings are transcribed for word timestamps automatically

Frontend UI improvements

VideoPanel.tsx - Two separate timeline bars: (1) Speech & Descriptions, (2) Gaps showing remaining available time after placed descriptions (full color = remaining, faded = used)

TranscriptPane.tsx - Dynamic gap reduction: gap blocks in the transcript visually shrink as descriptions are written into them

DescriptionCard.tsx
- Double-click time editing - double-click the start time to enter a numeric editor
- Quotation mark auto-locking - typing a description in "quotes" automatically sets locked_text
- Black/grey word colors - post-render, included words are black, excluded words are light grey (exact paper specification)
- Word selection syncs to server - toggling words or moving the slider immediately calls /word-selection and streams back re-cut audio

App.tsx
- TTS auto-preview during playback - when the video plays past a description's start time, the pre-generated TTS audio
plays automatically alongside the source audio
- Export All button - renders all three modes and shows audio players for each
\end{Verbatim}
\end{tcolorbox}

%% file: 02_x_bibliograph.bib
@inproceedings{yan2025lmr,
  title={Lmr-bench: Evaluating llm agent’s ability on reproducing language modeling research},
  author={Yan, Shuo and Li, Ruochen and Luo, Ziming and Wang, Zimu and Li, Daoyang and Jing, Liqiang and He, Kaiyu and Wu, Peilin and Ni, Juntong and Michalopoulos, George and others},
  booktitle={Proceedings of the 2025 Conference on Empirical Methods in Natural Language Processing},
  pages={6175--6197},
  year={2025}
}

@inproceedings{xcreation,
author = {Yan, Zihan and Yang, Chunxu and Liang, Qihao and Chen, Xiang 'Anthony'},
title = {XCreation: A Graph-based Crossmodal Generative Creativity Support Tool},
year = {2023},
isbn = {9798400701320},
publisher = {Association for Computing Machinery},
address = {New York, NY, USA},
url = {https://doi.org/10.1145/3586183.3606826},
doi = {10.1145/3586183.3606826},
booktitle = {Proceedings of the 36th Annual ACM Symposium on User Interface Software and Technology},
articleno = {48},
numpages = {15},
location = {San Francisco, CA, USA},
series = {UIST '23}
}

@inproceedings{worldsmith,
author = {Dang, Hai and Brudy, Frederik and Fitzmaurice, George and Anderson, Fraser},
title = {WorldSmith: Iterative and Expressive Prompting for World Building with a Generative AI},
year = {2023},
isbn = {9798400701320},
publisher = {Association for Computing Machinery},
address = {New York, NY, USA},
url = {https://doi.org/10.1145/3586183.3606772},
doi = {10.1145/3586183.3606772},
booktitle = {Proceedings of the 36th Annual ACM Symposium on User Interface Software and Technology},
articleno = {63},
numpages = {17},
location = {San Francisco, CA, USA},
series = {UIST '23}
}

@inproceedings{worldscribe,
author = {Chang, Ruei-Che and Liu, Yuxuan and Guo, Anhong},
title = {WorldScribe: Towards Context-Aware Live Visual Descriptions},
year = {2024},
isbn = {9798400706288},
publisher = {Association for Computing Machinery},
address = {New York, NY, USA},
url = {https://doi.org/10.1145/3654777.3676375},
doi = {10.1145/3654777.3676375},
booktitle = {Proceedings of the 37th Annual ACM Symposium on User Interface Software and Technology},
articleno = {140},
numpages = {18},
location = {Pittsburgh, PA, USA},
series = {UIST '24}
}

@inproceedings{waitgpt,
author = {Xie, Liwenhan and Zheng, Chengbo and Xia, Haijun and Qu, Huamin and Zhu-Tian, Chen},
title = {WaitGPT: Monitoring and Steering Conversational LLM Agent in Data Analysis with On-the-Fly Code Visualization},
year = {2024},
isbn = {9798400706288},
publisher = {Association for Computing Machinery},
address = {New York, NY, USA},
url = {https://doi.org/10.1145/3654777.3676374},
doi = {10.1145/3654777.3676374},
booktitle = {Proceedings of the 37th Annual ACM Symposium on User Interface Software and Technology},
articleno = {119},
numpages = {14},
location = {Pittsburgh, PA, USA},
series = {UIST '24}
}

@inproceedings{vizability,
author = {Gorniak, Joshua and Kim, Yoon and Wei, Donglai and Kim, Nam Wook},
title = {VizAbility: Enhancing Chart Accessibility with LLM-based Conversational Interaction},
year = {2024},
isbn = {9798400706288},
publisher = {Association for Computing Machinery},
address = {New York, NY, USA},
url = {https://doi.org/10.1145/3654777.3676414},
doi = {10.1145/3654777.3676414},
booktitle = {Proceedings of the 37th Annual ACM Symposium on User Interface Software and Technology},
articleno = {89},
numpages = {19},
location = {Pittsburgh, PA, USA},
series = {UIST '24}
}

@inproceedings{fogarty2017code,
  title={Code and contribution in interactive systems research},
  author={Fogarty, James},
  booktitle={Workshop HCITools: Strategies and Best Practices for Designing, Evaluating and Sharing Technical HCI Toolkits at CHI},
  pages={1--4},
  year={2017}
}

@inproceedings{ledo18,
author = {Ledo, David and Houben, Steven and Vermeulen, Jo and Marquardt, Nicolai and Oehlberg, Lora and Greenberg, Saul},
title = {Evaluation Strategies for HCI Toolkit Research},
year = {2018},
isbn = {9781450356206},
publisher = {Association for Computing Machinery},
address = {New York, NY, USA},
url = {https://doi.org/10.1145/3173574.3173610},
doi = {10.1145/3173574.3173610},
booktitle = {Proceedings of the 2018 CHI Conference on Human Factors in Computing Systems},
pages = {1–17},
numpages = {17},
location = {Montreal QC, Canada},
series = {CHI '18}
}

@inproceedings{greenberg08,
author = {Greenberg, Saul and Buxton, Bill},
title = {Usability evaluation considered harmful (some of the time)},
year = {2008},
isbn = {9781605580111},
publisher = {Association for Computing Machinery},
address = {New York, NY, USA},
url = {https://doi.org/10.1145/1357054.1357074},
doi = {10.1145/1357054.1357074},
booktitle = {Proceedings of the SIGCHI Conference on Human Factors in Computing Systems},
pages = {111–120},
numpages = {10},
location = {Florence, Italy},
series = {CHI '08}
}

@inproceedings{olsen07,
author = {Olsen, Dan R.},
title = {Evaluating user interface systems research},
year = {2007},
isbn = {9781595936790},
publisher = {Association for Computing Machinery},
address = {New York, NY, USA},
url = {https://doi.org/10.1145/1294211.1294256},
doi = {10.1145/1294211.1294256},
booktitle = {Proceedings of the 20th Annual ACM Symposium on User Interface Software and Technology},
pages = {251–258},
numpages = {8},
location = {Newport, Rhode Island, USA},
series = {UIST '07}
}

@inproceedings{validates,
author = {Shankar, Shreya and Zamfirescu-Pereira, J.D. and Hartmann, Bjoern and Parameswaran, Aditya and Arawjo, Ian},
title = {Who Validates the Validators? Aligning LLM-Assisted Evaluation of LLM Outputs with Human Preferences},
year = {2024},
isbn = {9798400706288},
publisher = {Association for Computing Machinery},
address = {New York, NY, USA},
url = {https://doi.org/10.1145/3654777.3676450},
doi = {10.1145/3654777.3676450},
booktitle = {Proceedings of the 37th Annual ACM Symposium on User Interface Software and Technology},
articleno = {131},
numpages = {14},
location = {Pittsburgh, PA, USA},
series = {UIST '24}
}

@inproceedings{synergi,
author = {Kang, Hyeonsu B and Wu, Tongshuang and Chang, Joseph Chee and Kittur, Aniket},
title = {Synergi: A Mixed-Initiative System for Scholarly Synthesis and Sensemaking},
year = {2023},
isbn = {9798400701320},
publisher = {Association for Computing Machinery},
address = {New York, NY, USA},
url = {https://doi.org/10.1145/3586183.3606759},
doi = {10.1145/3586183.3606759},
booktitle = {Proceedings of the 36th Annual ACM Symposium on User Interface Software and Technology},
articleno = {43},
numpages = {19},
location = {San Francisco, CA, USA},
series = {UIST '23}
}

@inproceedings{sketchnsketch,
author = {Hempel, Brian and Lubin, Justin and Chugh, Ravi},
title = {Sketch-n-Sketch: Output-Directed Programming for SVG},
year = {2019},
isbn = {9781450368162},
publisher = {Association for Computing Machinery},
address = {New York, NY, USA},
url = {https://doi.org/10.1145/3332165.3347925},
doi = {10.1145/3332165.3347925},
booktitle = {Proceedings of the 32nd Annual ACM Symposium on User Interface Software and Technology},
pages = {281–292},
numpages = {12},
location = {New Orleans, LA, USA},
series = {UIST '19}
}

@inproceedings{promptpaint,
author = {Chung, John Joon Young and Adar, Eytan},
title = {PromptPaint: Steering Text-to-Image Generation Through Paint Medium-like Interactions},
year = {2023},
isbn = {9798400701320},
publisher = {Association for Computing Machinery},
address = {New York, NY, USA},
url = {https://doi.org/10.1145/3586183.3606777},
doi = {10.1145/3586183.3606777},
booktitle = {Proceedings of the 36th Annual ACM Symposium on User Interface Software and Technology},
articleno = {6},
numpages = {17},
location = {San Francisco, CA, USA},
series = {UIST '23}
}

@ARTICLE{proactiveva,
  author={Zhao, Yuheng and Shu, Xueli and Fan, Liwen and Gao, Lin and Zhang, Yu and Chen, Siming},
  journal={IEEE Transactions on Visualization and Computer Graphics}, 
  title={ProactiveVA: Proactive Visual Analytics with LLM-Based UI Agent}, 
  year={2026},
  volume={32},
  number={1},
  pages={451-461},
  doi={10.1109/TVCG.2025.3642628}}

@inproceedings{patchview,
author = {Chung, John Joon Young and Kreminski, Max},
title = {Patchview: LLM-powered Worldbuilding with Generative Dust and Magnet Visualization},
year = {2024},
isbn = {9798400706288},
publisher = {Association for Computing Machinery},
address = {New York, NY, USA},
url = {https://doi.org/10.1145/3654777.3676352},
doi = {10.1145/3654777.3676352},
booktitle = {Proceedings of the 37th Annual ACM Symposium on User Interface Software and Technology},
articleno = {77},
numpages = {19},
location = {Pittsburgh, PA, USA},
series = {UIST '24}
}

@inproceedings{memorysandbox,
author = {Huang, Ziheng and Gutierrez, Sebastian and Kamana, Hemanth and Macneil, Stephen},
title = {Memory Sandbox: Transparent and Interactive Memory Management for Conversational Agents},
year = {2023},
isbn = {9798400700965},
publisher = {Association for Computing Machinery},
address = {New York, NY, USA},
url = {https://doi.org/10.1145/3586182.3615796},
doi = {10.1145/3586182.3615796},
booktitle = {Adjunct Proceedings of the 36th Annual ACM Symposium on User Interface Software and Technology},
articleno = {97},
numpages = {3},
location = {San Francisco, CA, USA},
series = {UIST '23 Adjunct}
}

@inproceedings{genassist,
author = {Huh, Mina and Peng, Yi-Hao and Pavel, Amy},
title = {GenAssist: Making Image Generation Accessible},
year = {2023},
isbn = {9798400701320},
publisher = {Association for Computing Machinery},
address = {New York, NY, USA},
url = {https://doi.org/10.1145/3586183.3606735},
doi = {10.1145/3586183.3606735},
booktitle = {Proceedings of the 36th Annual ACM Symposium on User Interface Software and Technology},
articleno = {38},
numpages = {17},
location = {San Francisco, CA, USA},
series = {UIST '23}
}

@inproceedings{coladder,
author = {Yen, Ryan and Zhu, Jiawen Stefanie and Suh, Sangho and Xia, Haijun and Zhao, Jian},
title = {CoLadder: Manipulating Code Generation via Multi-Level Blocks},
year = {2024},
isbn = {9798400706288},
publisher = {Association for Computing Machinery},
address = {New York, NY, USA},
url = {https://doi.org/10.1145/3654777.3676357},
doi = {10.1145/3654777.3676357},
booktitle = {Proceedings of the 37th Annual ACM Symposium on User Interface Software and Technology},
articleno = {11},
numpages = {20},
location = {Pittsburgh, PA, USA},
series = {UIST '24}
}

@inproceedings{cellsgens,
author = {Kim, Tae Soo and Lee, Yoonjoo and Chang, Minsuk and Kim, Juho},
title = {Cells, Generators, and Lenses: Design Framework for Object-Oriented Interaction with Large Language Models},
year = {2023},
isbn = {9798400701320},
publisher = {Association for Computing Machinery},
address = {New York, NY, USA},
url = {https://doi.org/10.1145/3586183.3606833},
doi = {10.1145/3586183.3606833},
booktitle = {Proceedings of the 36th Annual ACM Symposium on User Interface Software and Technology},
articleno = {4},
numpages = {18},
location = {San Francisco, CA, USA},
series = {UIST '23}
}

@article{xu2026scaling,
  title={Scaling Reproducibility: An AI-Assisted Workflow for Large-Scale Reanalysis},
  author={Xu, Yiqing and Yang, Leo Yang},
  journal={arXiv preprint arXiv:2602.16733},
  year={2026}
}

@article{seo2025paper2code,
  title={Paper2code: Automating code generation from scientific papers in machine learning},
  author={Seo, Minju and Baek, Jinheon and Lee, Seongyun and Hwang, Sung Ju},
  journal={arXiv preprint arXiv:2504.17192},
  year={2025}
}

@article{miao2025recode,
  title={Recode-h: A benchmark for research code development with interactive human feedback},
  author={Miao, Chunyu and Zou, Henry Peng and Li, Yangning and Chen, Yankai and Wang, Yibo and Wang, Fangxin and Li, Yifan and Yang, Wooseong and He, Bowei and Zhang, Xinni and others},
  journal={arXiv preprint arXiv:2510.06186},
  year={2025}
}

@article{jansen2025codedistiller,
  title={CodeDistiller: Automatically Generating Code Libraries for Scientific Coding Agents},
  author={Jansen, Peter and Hassan, Samiah and Narasimha, Pragnya},
  journal={arXiv preprint arXiv:2512.01089},
  year={2025}
}

@article{hua2025researchcodebench,
  title={Researchcodebench: Benchmarking llms on implementing novel machine learning research code},
  author={Hua, Tianyu and Hua, Harper and Xiang, Violet and Klieger, Benjamin and Truong, Sang T and Liang, Weixin and Sun, Fan-Yun and Haber, Nick},
  journal={arXiv preprint arXiv:2506.02314},
  year={2025}
}

@inproceedings{replichi13,
author = {Wilson, Max L. L. and Resnick, Paul and Coyle, David and Chi, Ed H.},
title = {RepliCHI: the workshop},
year = {2013},
isbn = {9781450319522},
publisher = {Association for Computing Machinery},
address = {New York, NY, USA},
url = {https://doi.org/10.1145/2468356.2479636},
doi = {10.1145/2468356.2479636},
booktitle = {CHI '13 Extended Abstracts on Human Factors in Computing Systems},
pages = {3159–3162},
numpages = {4},
location = {Paris, France},
series = {CHI EA '13}
}

@inproceedings{replichi11,
author = {Wilson, Max L. and Mackay, Wendy and Chi, Ed and Bernstein, Michael and Russell, Dan and Thimbleby, Harold},
title = {RepliCHI - CHI should be replicating and validating results more: discuss},
year = {2011},
isbn = {9781450302685},
publisher = {Association for Computing Machinery},
address = {New York, NY, USA},
url = {https://doi.org/10.1145/1979742.1979491},
doi = {10.1145/1979742.1979491},
booktitle = {CHI '11 Extended Abstracts on Human Factors in Computing Systems},
pages = {463–466},
numpages = {4},
location = {Vancouver, BC, Canada},
series = {CHI EA '11}
}

@inproceedings{replichi14,
author = {Wilson, Max L. and Chi, Ed H. and Reeves, Stuart and Coyle, David},
title = {RepliCHI: the workshop II},
year = {2014},
isbn = {9781450324748},
publisher = {Association for Computing Machinery},
address = {New York, NY, USA},
url = {https://doi.org/10.1145/2559206.2559233},
doi = {10.1145/2559206.2559233},
booktitle = {CHI '14 Extended Abstracts on Human Factors in Computing Systems},
pages = {33–36},
numpages = {4},
location = {Toronto, Ontario, Canada},
series = {CHI EA '14}
}

@inproceedings{replichi12,
author = {Wilson, Max and Mackay, Wendy and Chi, Ed and Bernstein, Michael and Nichols, Jeffrey},
title = {RepliCHI SIG: from a panel to a new submission venue for replication},
year = {2012},
isbn = {9781450310161},
publisher = {Association for Computing Machinery},
address = {New York, NY, USA},
url = {https://doi.org/10.1145/2212776.2212419},
doi = {10.1145/2212776.2212419},
booktitle = {CHI '12 Extended Abstracts on Human Factors in Computing Systems},
pages = {1185–1188},
numpages = {4},
location = {Austin, Texas, USA},
series = {CHI EA '12}
}

@article{siegel2024core,
  title={Core-bench: Fostering the credibility of published research through a computational reproducibility agent benchmark},
  author={Siegel, Zachary S and Kapoor, Sayash and Nagdir, Nitya and Stroebl, Benedikt and Narayanan, Arvind},
  journal={arXiv preprint arXiv:2409.11363},
  year={2024}
}

@article{starace2025paperbench,
  title={PaperBench: Evaluating AI's Ability to Replicate AI Research},
  author={Starace, Giulio and Jaffe, Oliver and Sherburn, Dane and Aung, James and Chan, Jun Shern and Maksin, Leon and Dias, Rachel and Mays, Evan and Kinsella, Benjamin and Thompson, Wyatt and others},
  journal={arXiv preprint arXiv:2504.01848},
  year={2025}
}

@article{luo2025executable,
  title={Executable Knowledge Graphs for Replicating AI Research},
  author={Luo, Yujie and Yu, Zhuoyun and Wang, Xuehai and Zhu, Yuqi and Zhang, Ningyu and Wei, Lanning and Du, Lun and Zheng, Da and Chen, Huajun},
  journal={arXiv preprint arXiv:2510.17795},
  year={2025}
}

@article{baumgartner2026scicoqa,
  title={SciCoQA: Quality Assurance for Scientific Paper--Code Alignment},
  author={Baumg{\"a}rtner, Tim and Gurevych, Iryna},
  journal={arXiv preprint arXiv:2601.12910},
  year={2026}
}

@inproceedings{Wacharamanotham20,
author = {Wacharamanotham, Chat and Eisenring, Lukas and Haroz, Steve and Echtler, Florian},
title = {Transparency of CHI Research Artifacts: Results of a Self-Reported Survey},
year = {2020},
isbn = {9781450367080},
publisher = {Association for Computing Machinery},
address = {New York, NY, USA},
url = {https://doi.org/10.1145/3313831.3376448},
doi = {10.1145/3313831.3376448},
booktitle = {Proceedings of the 2020 CHI Conference on Human Factors in Computing Systems},
pages = {1–14},
numpages = {14},
location = {Honolulu, HI, USA},
series = {CHI '20}
}

@article{cohoon2021norms,
  title={Norms and open systems in open science},
  author={Cohoon, Johanna and Howison, James},
  journal={Information \& Culture},
  volume={56},
  number={2},
  pages={115--137},
  year={2021},
  publisher={JSTOR}
}

@inproceedings{chen2025deep,
  title={Deep-Reproducer: From Paper Understanding to Code Generation},
  author={Chen, Pengcheng and Yan, Ning and Zhao, Zihan and Lin, Yixiao and Chen, Huaibo and Hu, Yue and Bai, Qinbo and Li, Xiang and Mortazavi, Masood S},
  booktitle={NeurIPS 2025 Fourth Workshop on Deep Learning for Code},
  year = {2025}
}

@inproceedings{Echtler18,
author = {Echtler, Florian and H\"{a}u\ss{}ler, Maximilian},
title = {Open Source, Open Science, and the Replication Crisis in HCI},
year = {2018},
isbn = {9781450356213},
publisher = {Association for Computing Machinery},
address = {New York, NY, USA},
url = {https://doi.org/10.1145/3170427.3188395},
doi = {10.1145/3170427.3188395},
booktitle = {Extended Abstracts of the 2018 CHI Conference on Human Factors in Computing Systems},
pages = {1–8},
numpages = {8},
location = {Montreal QC, Canada},
series = {CHI EA '18}
}

@inproceedings{isenberg2024state,
  title={The state of reproducibility stamps for visualization research papers},
  author={Isenberg, Tobias},
  booktitle={2024 IEEE Evaluation and Beyond-Methodological Approaches for Visualization (BELIV)},
  pages={97--105},
  year={2024},
  organization={IEEE}
}

@article{zhao2025autoreproduce,
  title={Autoreproduce: Automatic ai experiment reproduction with paper lineage},
  author={Zhao, Xuanle and Sang, Zilin and Li, Yuxuan and Shi, Qi and Zhao, Weilun and Wang, Shuo and Zhang, Duzhen and Han, Xu and Liu, Zhiyuan and Sun, Maosong},
  journal={arXiv preprint arXiv:2505.20662},
  year={2025}
}

@article{majrashi2023inter,
  title={Inter-platform consistency inspection method},
  author={Majrashi, Khalid},
  journal={International Journal of Technology and Human Interaction (IJTHI)},
  volume={19},
  number={1},
  pages={1--20},
  year={2023},
  publisher={IGI Global Scientific Publishing}
}

@article{dix2010human,
  title={Human--computer interaction: A stable discipline, a nascent science, and the growth of the long tail},
  author={Dix, Alan},
  journal={Interacting with computers},
  volume={22},
  number={1},
  pages={13--27},
  year={2010},
  publisher={OUP}
}

@inproceedings{hu2025repro,
  title={REPRO-BENCH: Can Agentic AI Systems Assess the Reproducibility of Social Science Research?},
  author={Hu, Chuxuan and Zhang, Liyun and Lim, Yeji and Wadhwani, Aum and Peters, Austin and Kang, Daniel},
  booktitle={Findings of the Association for Computational Linguistics: ACL 2025},
  pages={23616--23626},
  year={2025}
}

@inproceedings{belen2025seems,
  title={"Seems Complicated and Unachievable": A Collaborative Autoethnography of Design Replication in Undergraduate Research},
  author={Belen Saavedra Rios, Maria and Takei, Youngsoon and Hicks, Blade and Stamato, Lydia and Wu, Wei and Jones, Jasmine},
  booktitle={Proceedings of the 7th Annual Symposium on HCI Education},
  pages={1--11},
  year={2025}
}

@article{si2025ideation,
  title={The ideation-execution gap: Execution outcomes of llm-generated versus human research ideas},
  author={Si, Chenglei and Hashimoto, Tatsunori and Yang, Diyi},
  journal={arXiv preprint arXiv:2506.20803},
  year={2025}
}

@article{edwards2025rexbench,
  title={RExBench: Can coding agents autonomously implement AI research extensions?},
  author={Edwards, Nicholas and Lee, Yukyung and Mao, Yujun Audrey and Qin, Yulu and Schuster, Sebastian and Kim, Najoung},
  journal={arXiv preprint arXiv:2506.22598},
  year={2025}
}

@inproceedings{hornbaek2014once,
  title={Is once enough? On the extent and content of replications in human-computer interaction},
  author={Hornb{\ae}k, Kasper and Sander, S{\o}ren S and Bargas-Avila, Javier Andr{\'e}s and Grue Simonsen, Jakob},
  booktitle={Proceedings of the SIGCHI conference on human factors in computing systems},
  pages={3523--3532},
  year={2014}
}

@inproceedings{salehzadeh2023changes,
  title={Changes in research ethics, openness, and transparency in empirical studies between CHI 2017 and CHI 2022},
  author={Salehzadeh Niksirat, Kavous and Goswami, Lahari and SB Rao, Pooja and Tyler, James and Silacci, Alessandro and Aliyu, Sadiq and Aebli, Annika and Wacharamanotham, Chat and Cherubini, Mauro},
  booktitle={Proceedings of the 2023 CHI conference on human factors in computing systems},
  pages={1--23},
  year={2023}
}

@inproceedings{oppenlaender2025keeping,
  title={Keeping score: A quantitative analysis of how the CHI community appreciates its milestones},
  author={Oppenlaender, Jonas and Hosio, Simo},
  booktitle={Proceedings of the 2025 CHI Conference on Human Factors in Computing Systems},
  pages={1--17},
  year={2025}
}

@inproceedings{vanderdonckt2025context,
  title={Context is Key for Reproducibility of Empirical Studies in Human-Computer Interaction},
  author={Vanderdonckt, Jean and Vatavu, Radu-Daniel},
  booktitle={Proceedings of the 3rd ACM Conference on Reproducibility and Replicability},
  pages={41--50},
  year={2025}
}

@article{ge2025survey,
  title={A survey of vibe coding with large language models},
  author={Ge, Yuyao and Mei, Lingrui and Duan, Zenghao and Li, Tianhao and Zheng, Yujia and Wang, Yiwei and Wang, Lexin and Yao, Jiayu and Liu, Tianyu and Cai, Yujun and others},
  journal={arXiv preprint arXiv:2510.12399},
  year={2025}
}

@article{syeda2024vis,
  title={Vis repligogy: Towards a culture of facilitating replication studies in visualization pedagogy and research},
  author={Syeda, Uzma Haque and South, Laura and Raynor, Justin and Panavas, Liudas and Saffo, David and Morriss, Tommy and Dunne, Cody and Borkin, Michelle A},
  year={2024},
  publisher={The Eurographics Association}
}

@misc{acmbadge,
title={Artifact Review and Badging - Current},
lastaccessed={March 21, 2026},
url={https://www.acm.org/publications/policies/artifact-review-and-badging-current},
year={2020},
author={{Association for Computing Machinery}}
}

@misc{whenwords,
title={whenwords: An Open Source Library Without Code},
lastacessed={March 16, 2026},
author={Drew Breunig},
url={https://github.com/dbreunig/whenwords},
year={2026}
}

@misc{vibedef,
author={Simon Willison},
date={March 19, 2025},
title={Not all AI-assisted programming is vibe coding (but vibe coding rocks)},
lastaccessed={March 14, 2026},
url={https://simonwillison.net/ 2025/Mar/19/vibe-coding/}
}

@inproceedings{lau2025design,
  title={The Design Space of LLM-Based AI Coding Assistants: An Analysis of 90 Systems in Academia and Industry},
  author={Lau, Sam and Guo, Philip J},
  booktitle={2025 IEEE Symposium on Visual Languages and Human-Centric Computing (VL/HCC)},
  pages={300--313},
  year={2025},
  organization={IEEE}
}

@misc{antigravitybrowser,
author={{Google Antigravity}},
title={Browser},
year={2026},
lastaccessed={March 14, 2026},
url={https://antigravity.google/docs/browser}
}

@misc{cursorbrowser,
author={{Cursor}},
title={Browser},
year={2026},
lastaccessed={March 14, 2026},
url={https://cursor.com/docs/agent/tools/browser}
}

@article{ye2025replicationbench,
  title={ReplicationBench: Can AI Agents Replicate Astrophysics Research Papers?},
  author={Ye, Christine and Yuan, Sihan and Cooray, Suchetha and Dillmann, Steven and Roque, Ian LV and Baron, Dalya and Frank, Philipp and Martin-Alvarez, Sergio and Koblischke, Nolan and Qu, Frank J and others},
  journal={arXiv preprint arXiv:2510.24591},
  year={2025}
}

@article{ding2025nl2repo,
  title={NL2Repo-Bench: Towards Long-Horizon Repository Generation Evaluation of Coding Agents},
  author={Ding, Jingzhe and Long, Shengda and Pu, Changxin and Zhou, Huan and Gao, Hongwan and Gao, Xiang and He, Chao and Hou, Yue and Hu, Fei and Li, Zhaojian and others},
  journal={arXiv preprint arXiv:2512.12730},
  year={2025}
}

@inproceedings{Ballou2021,
author = {Ballou, Nick and Warriar, Vivek R. and Deterding, Sebastian},
title = {Are You Open? A Content Analysis of Transparency and Openness Guidelines in HCI Journals},
year = {2021},
isbn = {9781450380966},
publisher = {Association for Computing Machinery},
address = {New York, NY, USA},
url = {https://doi.org/10.1145/3411764.3445584},
doi = {10.1145/3411764.3445584},
booktitle = {Proceedings of the 2021 CHI Conference on Human Factors in Computing Systems},
articleno = {176},
numpages = {10},
location = {Yokohama, Japan},
series = {CHI '21}
}

@inproceedings{b2,
author = {Wu, Yifan and Hellerstein, Joseph M. and Satyanarayan, Arvind},
title = {B2: Bridging Code and Interactive Visualization in Computational Notebooks},
year = {2020},
isbn = {9781450375146},
publisher = {Association for Computing Machinery},
address = {New York, NY, USA},
url = {https://doi.org/10.1145/3379337.3415851},
doi = {10.1145/3379337.3415851},
booktitle = {Proceedings of the 33rd Annual ACM Symposium on User Interface Software and Technology},
pages = {152–165},
numpages = {14},
location = {Virtual Event, USA},
series = {UIST '20}
}

@inproceedings{qassem25,
author = {Qassem, Asrar and Bryce, Renee and Alkhaldi, Khalid},
year = {2025},
month = {09},
pages = {108-113},
title = {A Survey on Visual GUI Testing for Automatic Test Case Generation: Tools, AI Techniques, and Emerging Trends (S)},
doi = {10.18293/SEKE2025-108}
}

@article{ramler2018adapting,
  title={Adapting automated test generation to GUI testing of industry applications},
  author={Ramler, Rudolf and Buchgeher, Georg and Klammer, Claus},
  journal={Information and Software Technology},
  volume={93},
  pages={248--263},
  year={2018},
  publisher={Elsevier}
}

@book{hopkins2008eating,
  title={Eating the IT elephant: Moving from greenfield development to brownfield},
  author={Hopkins, Richard and Jenkins, Kevin},
  year={2008},
  publisher={Addison-Wesley Professional}
}

@inproceedings{rescribe,
author = {Pavel, Amy and Reyes, Gabriel and Bigham, Jeffrey P.},
title = {Rescribe: Authoring and Automatically Editing Audio Descriptions},
year = {2020},
isbn = {9781450375146},
publisher = {Association for Computing Machinery},
address = {New York, NY, USA},
url = {https://doi.org/10.1145/3379337.3415864},
doi = {10.1145/3379337.3415864},
booktitle = {Proceedings of the 33rd Annual ACM Symposium on User Interface Software and Technology},
pages = {747–759},
numpages = {13},
location = {Virtual Event, USA},
series = {UIST '20}
}

@inproceedings{mage,
author = {Kery, Mary Beth and Ren, Donghao and Hohman, Fred and Moritz, Dominik and Wongsuphasawat, Kanit and Patel, Kayur},
title = {mage: Fluid Moves Between Code and Graphical Work in Computational Notebooks},
year = {2020},
isbn = {9781450375146},
publisher = {Association for Computing Machinery},
address = {New York, NY, USA},
url = {https://doi.org/10.1145/3379337.3415842},
doi = {10.1145/3379337.3415842},
booktitle = {Proceedings of the 33rd Annual ACM Symposium on User Interface Software and Technology},
pages = {140–151},
numpages = {12},
location = {Virtual Event, USA},
series = {UIST '20}
}

@inproceedings{spellburst,
author = {Angert, Tyler and Suzara, Miroslav and Han, Jenny and Pondoc, Christopher and Subramonyam, Hariharan},
title = {Spellburst: A Node-based Interface for Exploratory Creative Coding with Natural Language Prompts},
year = {2023},
isbn = {9798400701320},
publisher = {Association for Computing Machinery},
address = {New York, NY, USA},
url = {https://doi.org/10.1145/3586183.3606719},
doi = {10.1145/3586183.3606719},
booktitle = {Proceedings of the 36th Annual ACM Symposium on User Interface Software and Technology},
articleno = {100},
numpages = {22},
location = {San Francisco, CA, USA},
series = {UIST '23}
}
